\documentstyle[aps,eqsecnum,epsf,preprint,tighten]{revtex}
\newcommand{\bk}{{\bf k}}
\newcommand{\bx}{{\bf x}}
\newcommand{\bp}{{\bf p}}
\newcommand{\bq}{{\bf q}}
\newcommand{\bpsi}{{\overline{\psi}}}
\begin{document}
\widetext
\title{Non-zero temperature transport near\\
fractional quantum Hall critical points}
\author{Subir Sachdev}
\address{
Department of Physics, P.O. Box 208120,
Yale University, New Haven, CT 06520-8120}
\date{\today}
\maketitle
\begin{abstract}
In an earlier work, Damle and the author (Phys. Rev. B in press;
cond-mat/9705206) demonstrated
the central role played by incoherent, inelastic processes in transport
near two-dimensional quantum critical points. This paper extends these results
to the case of a quantum transition in an anyon gas
between a fractional quantized Hall state and an insulator,
induced by varying the strength of an external periodic
potential. We use the quantum field theory for this transition introduced by
Chen, Fisher and Wu (Phys. Rev. B {\bf 48}, 13749 (1993)). The longitudinal and Hall
conductivities at the critical point are both $e^2/ h$ times non-trivial, fully
universal functions of
$\hbar \omega / k_B T$ ($\omega$ is the measuring frequency). These functions
are computed using a combination of perturbation theory
on the Kubo formula, and the solution of a quantum Boltzmann equation for
the anyonic quasiparticles and quasiholes.
The results include the values
of the d.c. conductivities ($\hbar \omega /k_B T \rightarrow 0$); earlier work
had been restricted strictly to $T=0$, and had therefore
computed only the high frequency a.c. conductivities with $\hbar \omega / k_B T \rightarrow 
\infty$. 
\end{abstract}
\pacs{PACS numbers:}
\widetext
\newpage
\section{Introduction}
\label{intro}
The remarkably rich zero temperature phase diagram of a two-dimensional electron
gas in a high magnetic field~\cite{sankar} offers an ideal laboratory for exploring the
physics of systems near a quantum critical point. Much theoretical and experimental
attention has focussed on the transitions between 
the various quantum Hall plateau states and the insulator~\cite{bodo,shahar}. 

The vast majority of the theoretical work has studied transitions between integer
quantum Hall plateaus using a model of non-interacting electrons undergoing
phase-coherent transport in the presence of a random potential~\cite{bodo,sankar1}: this 
transport can be described by non-linear sigma model field theories~\cite{llp}
or by a network model~\cite{cc}. In this scenario, interactions only
provide a phase-breaking length which cuts off the critical quantum interference effects,
but are irrelevant perturbations at the quantum-critical point.
It has also been suggested that the transitions
involving fractional quantum Hall plateaus can be described using closely 
related models of non-interacting composite fermions~\cite{jkt,klz,clee}.

However, there have been a number of indications in recent experiments
that interactions play a more fundamental role~\cite{shahar,ps,ds}.
Measurements of the non-linear transport~\cite{wei,shahar} have allowed direction measurement
of the dynamic critical exponent $z$: the measured value $z \approx 1$ is
incompatible  with critical points of non-interacting electrons, but can possibly
be understood to be a consequence of the Coulomb repulsion at an interacting
critical point~\cite{fgg,ps}. A more direct indication of the importance
comes from measurements of frequency dependent transport.
It was found that the width of transition region between the plateaus scales with
the same power of frequency, $\omega$, and temperature, $T$~\cite{engel}: 
such a property only holds
at quantum transitions described by a fixed point with non-zero interactions, while
non-interacting fixed points with irrelevant interactions have very different sensitivities
to perturbations in $\omega$ and $T$~\cite{sro}.

Non-zero temperature 
transport near interacting quantum critical points in 
two dimensions was studied by Damle and author in a previous paper~\cite{ds}
(hereafter referred to as I). 
A number of general observations were made, supported by
an explicit calculation 
on the superfluid-insulator transition in a Mott-Hubbard model for bosons.
In particular, it was argued that  the 
conductivity at the critical point 
was given by $e^2 /h$ times a non-trivial universal function
of the ratio, $\hbar \omega / k_B T$. Attention was drawn
to the rather different physical properties of the limiting regimes 
of frequency and temperature: ({\em i\/}) $\hbar \omega \gg k_B T$
with phase coherent transport of excitations created by the external
electric field, and ({\em ii\/}) $\hbar \omega \ll k_B T$, with
incoherent, quantum relaxational transport involving repeated inelastic
scattering between pre-existing thermally, excited carriers, and relaxation
to local equilibrium. The d.c. transport
is clearly in the latter regime, and it was noted in I that the d.c.
conductivity was universal even though it was dominated by inelastic
processes: this was possible because the cross-section for scattering between
the carriers became universal at the quantum critical point.

This paper will extend the considerations of I to one of the 
simplest possible {\em interacting\/} quantum critical points in a fractional quantum Hall
system. Any experimentally realistic model must account for both the
Coulomb interactions among the electrons and for the external random potential.
Such a model is of forbidding complexity, and we shall choose here to 
consider non-random models in the presence of an external {\em periodic}
potential with only a statistical gauge force interaction between the excitations. 
Our choice is motivated by the desire to explore seriously the consequences
of interactions in the simplest possible context, 
and the arguments above that inelastic, incoherent
processes are essential to any analysis of the experimentally observed critical
properties. 
The absence of a random potential means that there are no localized states:
these are essential for the appearance of quantum Hall plateaus as a function of the
external magnetic field, as they appear when the Fermi level is in a region of localized
states. However, the localized states are not expected to be as crucial in the 
experimentally relevant critical region (the studies of non-interacting 
electrons in a random potential examine the complicated
physics of the divergence of the localization length at the critical point in
perfectly phase coherent transport, but this is always cut off in interacting
models at some finite scale
due to incoherent, inelastic processes present 
at any non-zero temperature).
A second motivation for examining the case of the periodic potential comes from
its importance in the early studies of the integer quantum Hall effect~\cite{tknn}:
the topological nature of the quantization of the Hall conductivity first
emerged in this context.

A number of earlier studies have examined the fractional quantum Hall effect in a 
periodic potential. Kol and Read~\cite{kol} examined general global criteria that
must be satisfied by such states. Wen and Wu~\cite{ww} and Chen, Fisher and Wu~\cite{cfw} 
first examined the quantum critical points between different 
fractional quantum Hall states
and the insulator, and developed a low energy 
field-theoretic descriptions of the transitions.
They argued that these transitions could be understood in terms of a Mott transition
of an anyon gas in a periodic potential and an external magnetic field. The field theory
of the anyon gas consisted of a $U(1)$ Chern-Simons gauge field coupled to either relativistic
bosonic or Dirac Fermi fields. In an interesting recent work, Pfannkuche and 
MacDonald~\cite{allan} have numerically shown the presence of a transition between a fractional 
quantum Hall state and an insulator induced by varying the strength of an external periodic 
potential acting on an electron gas in a strong magnetic field: they also present evidence
for scaling behavior in the spectrum of small systems near the critical point.

This paper shall describe the nature of non-zero $T$ and $\omega$ transport
in the $U(1)$ Chern-Simons gauge field coupled to Dirac Fermi field  model
of Chen, Fisher and Wu~\cite{cfw}. Our motivation for chosing this particular model is
mainly technical:  it provides the simplest possible context for describing
the physics of incoherent transport in quantum Hall critical points, and allows
us to keep technical details to a minimum. Our results can also be extended to
models of bosons coupled to a Chern-Simons gauge field using a $1/N$ expansion,
and this will be discussed in Appendix~\ref{largeN}. 
We will show that at the critical point, the longitudinal ($\sigma_{xx}$) and Hall 
($\sigma_{xy}$) conductivities obey the scaling forms
\begin{equation}
\sigma_{xx} = \frac{e^2}{h} \Sigma_{xx} \left( \frac{\hbar\omega}{k_B T} \right)~~~~~;~~~~~
\sigma_{xy} = \frac{e^2}{h} \Sigma_{xy} \left( \frac{\hbar\omega}{k_B T} \right)
\label{scale}
\end{equation}
We shall extend the quantum transport theory of I to obtain explicit expressions
for the universal functions $\Sigma_{xx}$ and $\Sigma_{xy}$. The earlier
computations of these conductivities~\cite{cfw} were carried out precisely
at $T=0$, and therefore yielded the values of $\Sigma_{xx} ( \infty )$ and
$\Sigma_{xy} (\infty)$ which describe phase-coherent transport. 
The d.c. conductivities are however given by $\Sigma_{xx} (0)$
and $\Sigma_{xy} ( 0)$: these are quite different, but also universal numbers,
characterizing incoherent, relaxational transport,
and will be computed in this paper. 

As many of the considerations of this paper are similar to those of I,
we now explicitly point out some aspects which are different.
The central focus of our analysis shall not be the conductivities $\sigma$, but
related irreducible conductivities $\widetilde{\sigma}$: these are diagrammatically
defined as being irreducible towards cutting any single Chern-Simons gauge field propagator.
We shall find that many of the issues arising in the computation of $\widetilde{\sigma}_{xx}$
are similar to those discussed in I: the focus shall be on a quasiparticle-quasihole
contribution which is dominated by collisions among the excitations,
and is best described by a quantum transport equation. The physics going into the 
computation of $\widetilde{\sigma}_{xy}$ is quite different. The real part of 
$\widetilde{\sigma}_{xy}$ is non-dissipative and is actually best computed by
a simple perturbation theory, which already yields a result which is an 
interesting function of $\hbar \omega / k_B T$.  Indeed, such a perturbation theory
was presented in Ref~\cite{cfw} at $T=0$, and here we consider the orders of 
limit of $\omega \rightarrow 0$ and $T \rightarrow 0$ more carefully.
Our final results for $\sigma_{xx}$ also has a form rather different
from that obtained in I: it is given by complicated interplay between
different contributions to $\widetilde{\sigma}$: in particular we find that in the
low frequency collision-dominated regime it is possible to have $\sigma_{xx}$ be
an increasing, rather than decreasing, function of $\hbar \omega / k_B T$.
For the latter situation we find that $\sigma_{xx} ( \omega ) $ has a peak
at $\omega \sim k_B T/ \hbar$, and eventually decreases for large enough $\omega$.

As noted above, an essential ingredient in our computation is the quantum transport
equation for quasiparticles and quasiholes. These excitations are anyons, and so their
scattering cross-section is singular in the forward direction: for momentum
transfer $q$, the cross-section $\sim 1/q^2$~\cite{wilczek}. This scattering cross-section
appears in the collision term of the transport equation, and could potentially cause
infrared divergences. However we shall show
explicitly that the potential divergences all cancel, and the physical transport
co-efficients of the anyon gas are well defined. This is an important, albeit
technical, consistency check of our analysis, and occupies a significant part of
Section~\ref{qtrans}.

The outline of the remainder of the paper is as follows. We will introduce the model
and discuss some important general issues in Section~\ref{model}.
The perturbation theory for $\widetilde{\sigma}$ will be presented in Section~\ref{kubo}:
this will allow us to identify the different physical contributions to $\widetilde{\sigma}$.
One of these, the quasiparticle/quasihole contribution is a delta function
at zero frequency at leading order in $\alpha$. This is an artifact, and
has to be repaired by an infinite order resummation: the latter is most
conveniently carried out using a quantum transport equation analysis which
will be discussed in Section~\ref{qtrans}. We will synthesize our results for
$\widetilde{\sigma}$ into those for the physically observable $\sigma$ in Section~\ref{synth},
and conclude with some general remarks in Section~\ref{conc}.

\section{The Model and some general considerations}
\label{model}

To the extent possible, we will follow the notation of Ref~\onlinecite{cfw}.
We will be working the relativistic models with a velocity $c$, and 
use units with $\hbar = c = k_B = 1$, except where we explicitly quote conductivities
in units of $e^2 / h$.
The imaginary time action 
of Dirac fermions coupled to a $U(1)$ Chern-Simons gauge field introduced in
Ref~\onlinecite{cfw} is 
\begin{eqnarray}
{\cal S} = &&  \int_0^{1/T} d \tau \int d \bx \biggl\{
\bpsi_1 \left[ \gamma_{\mu} \left( 
\partial_{\mu} - i g a_{\mu} - i qe A_{\mu} \right) + i M_1 \right] \psi_1 \nonumber \\
&&~~~~~~~~~~~~~~~~~~~~~~~~~\left. + \bpsi_2 \left[ \gamma_{\mu} \left( 
\partial_{\mu} - i g a_{\mu} - i qe A_{\mu} \right) + i M_2 \right] \psi_2 + \frac{i}{2} 
\epsilon_{\mu\nu\lambda} a_{\mu} \partial_{\nu} a_{\lambda} \right\}
\label{cs1}
\end{eqnarray}
where  $\bx \equiv (x,y)$, and the Greek indices extend over $x,y,\tau$.
The Chern-Simons gauge field is $a_{\mu}$, the external vector potential
driving the currents in $A_{\mu}$, and $\psi_{1,2}$ are two species of 
Dirac fermions with charge $qe$. 
The Dirac matrices are $\gamma_{\mu} = i (- \tau_y, \tau_x, \tau_z)$ 
 where $\tau$ are the Pauli matrices, and $\bpsi = \psi^{\dagger}
\gamma_{\tau}$.
Notice that we are using the symbol $\tau$ to indicate both imaginary
time and the Pauli matrices, but the meaning should be clear from the context.
The sole role of $a_{\mu}$ and the Chern-Simons term
is to attach flux tubes onto each Dirac fermion and to convert it to
a Dirac anyon with statistics parameter $(1-\alpha)$ where
\begin{equation}
\alpha = \frac{g^2}{ 2 \pi};
\end{equation}
we are using the nomenclature where a statistics parameter of 1 corresponds
to fermions, and 0 to bosons. Notice that we have two species of anyons, $\psi_{1,2}$:
these are required by the doubling theorems for chiral fermions on a lattice. 
The anyons/anti-anyons represent the Laughlin quasiparticles/quasiholes 
of the fractional quantized Hall state. So to describe physics in the vicinity
of the simplest Laughlin quantum Hall state at filling $\nu=1/3$ we should choose
$\alpha = 2/3$ and $q=1/3$. As we will review below, the action ${\cal S}$
describes~\cite{cfw} a quantum transition between fractional quantized Hall state
with $T=0$ conductivity (the Latin indices extend over $x,y$)
\begin{equation}
\sigma_{ij} = -\epsilon_{ij} \frac{e^2}{h} \frac{q^2}{(1-\alpha)}
\label{fqh}      
\end{equation}
and an insulator with $\sigma_{ij} = 0$. Notice that, as expected, (\ref{fqh}) gives a Hall
conductance of $e^2 / (3 h)$ for $q=1/3$ and $\alpha = 2/3$.
 
We now make some general remarks about the nature of charge transport
in Chern-Simons theories: these results are well known~\cite{klz,cfw}, but our
formulation is especially suited to our subsequent analysis of inelastic
processes using the quantum Boltzmann equation.

From the equation of motion of $a_{\mu}$ implied by ${\cal S}$, we see that
role of $a_{\mu}$ is to impose the constraint
\begin{equation}
\epsilon_{\mu\nu\lambda} \partial_{\nu} a_{\lambda} = g 
\left( \bpsi_1 \gamma_{\mu} \psi_1 + \bpsi_2 \gamma_{\mu} \psi_2 \right).
\label{cs2}
\end{equation}
This constraint will be imposed {\em exactly\/} in all stages of our analysis.
The $\tau$ component of this constraint is 
\begin{equation}
\epsilon_{ij} \partial_i a_j = - g \left( \psi_1^{\dagger} \psi_1 + \psi_2^{\dagger}
\psi_2 \right).
\label{cs3}
\end{equation}
where the Latin indices extend over $x,y$.
The spatial components of (\ref{cs2}) can be written as
\begin{equation}
-qe \epsilon_{ij} E_{aj} = g \left(J_{1i} + J_{2i} \right)
\label{cs4}
\end{equation}
where ${\bf E}_a$ is the `electric field' associated with the $a_{\mu}$ 
gauge field
\begin{equation}
E_{ai} =  \partial_{\tau} a_{i} - \partial_i a_{\tau},
\label{cs5}
\end{equation}
and ${\bf J}_1$, ${\bf J}_2$ are the electrical currents carried by the
two species of Dirac fermions:
\begin{equation}
J_{1i} = qe \bpsi_1 \gamma_i \psi_1
\label{cs6}
\end{equation}
and similarly for ${\bf J}_2$.

We shall be interested in the response of ${\cal S}$ to an external physical electric
field ${\bf E}$ given, of course, by
\begin{equation}
E_{i} =  \partial_{\tau} A_{i} - \partial_i A_{\tau},
\label{cs7}
\end{equation}
The currents induced by such a field, in turn, induce a non-zero value of the
${\bf E}_a$ field, by (\ref{cs4}). It is therefore useful to define 
\begin{equation}
{\bf E}_T = {\bf E} + \frac{g}{qe} {\bf E}_a
\label{cs8}
\end{equation}
as the {\em total} effective field experienced by the Dirac fermions.
The field ${\bf E}_T$ will appear naturally in our subsequent formulation
of the quantum transport equations.
It is then useful to consider the response of the fermions by introducing the
linear response relations
\begin{equation}
J_{1i} ( \omega ) = \widetilde{\sigma}_{1ij} ( \omega ) E_{Tj} ( \omega ) 
\label{cs9}
\end{equation}
and similarly for ${\bf J}_2$. We have also introduced a frequency
$\omega$ to allow for the response to a time-dependent electric field.
In diagrammatic terms, if we compute $\widetilde{\sigma}_{\pm}$ via a Kubo
formula, we should only include diagrams which are irreducible towards
cutting any $a_{\mu}$ propagator: the irreducibility requirement
is equivalent to considering the response to the effective field
${\bf E}_T$.

The experimentally measured response, $\sigma_{ij}$, of course,
relates the total current to the  
external electric field
\begin{equation}
J_{1i} (\omega ) + J_{2 i } ( \omega ) =
\sigma_{ij} ( \omega ) 
E_j ( \omega )
\label{cs10}
\end{equation}
From (\ref{cs4},\ref{cs8},\ref{cs9},\ref{cs10}) we can obtain the following
basic relationship~\cite{klz,cfw} between the response functions
\begin{equation}
(\sigma^{-1})_{ij} = ( \widetilde{\sigma}_1 + \widetilde{\sigma}_2 
)^{-1}_{ij} - \frac{h}{q^2 e^2}
\epsilon_{ij} \alpha.
\label{cs11}
\end{equation}

When we are well away from the critical point, $|M_{1,2}|$ is large, 
and the $\sigma_{1,2}$ are easily computed at $T=0$; their exact values
are~\cite{cfw}
\begin{equation}
\widetilde{\sigma}_{1,2} = - \frac{q^2 e^2}{2 h} \epsilon_{ij} \mbox{sgn} (M_{1,2}).
\label{cs12}
\end{equation}
We see therefore that with  the choices $M_{1} > 0$ and $M_{2} > 0$,
(\ref{cs11},\ref{cs12}) lead to the conductivity (\ref{fqh}) of the fractional
quantized Hall state. The system undergoes a transition to an insulator
if $M_{1}$ changes sign while $M_{2}$ remains positive: then the total
conductivity vanishes. 

The remainder of the paper will study the critical point
where $M_{2}$ is large and positive (and therefore the
$\psi_2$ fermionic excitations have large gap) while
the renormalized $M_{1} = 0$ at $T=0$. 
The $\widetilde{\sigma}_{2}$ conductivity therefore continues
to satisfy (\ref{cs12}) at all low $T$ and $\omega$; all corrections
are suppressed by factors of $e^{-M_2 /T}$.
The response of the $\psi_1$ fermions is much more non-trivial, and will 
occupy the remainder of the paper to elucidate: it has a highly 
non-trivial, universal function of $\omega /T$.
As our attention will focus solely on the response of the $\psi_1$
fermions, we will henceforth drop the subscript $1$, and simply
refer to them as the $\psi$ fermions.

The central purpose of this paper is to analyze the quantum transport 
properties of the model ${\cal S}$ in (\ref{cs1}) at $T>0$. While the action ${\cal S}$
is Lorentz invariant, the presence of a nonzero $T$ breaks the Lorentz
symmetry, and the space and time dependence of observable correlators will
be quite different. Therefore, there is no particular advantage to working
with a covariant gauge, and it pays to choose a gauge better suited to physics
at $T>0$. It turns out that by far the most convenient choice is the Coulomb gauge:
the perturbative calculations to be considered in Section~\ref{kubo} are
the most straightforward, and the quantum transport equations
of Section~\ref{qtrans} are readily derived in this gauge. The Coulomb gauge 
corresponds to the choice
\begin{equation}
\partial_i a_i = 0,
\end{equation} 
combined with an exact integral of ${\cal S}$ over $a_{\tau}$.
Then the constraint (\ref{cs3}) can be solved to give
\begin{equation}
a_i = g \frac{\epsilon_{ij} \partial_j}{\partial^2} \psi^{\dagger} \psi
\label{cs14}
\end{equation}
where $\partial^2 \equiv \partial_i \partial_i$.
Note that this is an equal-time relationship--the absence of retardation in the
gauge field interactions is an important simplifying feature of this gauge.
Inserting (\ref{cs14}) into (\ref{cs1}) (and dropping the non-critical $\psi_2$ fermions)
we obtain the Hamiltonian density
\begin{eqnarray}
H &=& H_0 + H_1 \nonumber \\
H_0 &=& \int d \bx \left[
 \psi^{\dagger} \left( -i\tau_i \partial_i - M_0 \tau_z \right) \psi \right] \\
H_1 &=& \int d \bx \left[
2 \pi \alpha \psi^{\dagger} \epsilon_{ij}\tau_j \psi \frac{\partial_i}{\partial^2}
\psi^{\dagger} \psi \right].
\end{eqnarray}
We have inserted a {\em bare} mass $M_0$ in $H_0$; to ensure that the system is
at its critical point, the value of $M_0$ will have to be adjusted
order by order in $\alpha$ so that the {\em renormalized} mass vanishes at $T=0$. 
Note that the gauge field has disappeared as an independent dynamical degree of freedom:
each Dirac particle/anti-particle now has a flux tube attached, and it is this
flux tube which is responsible for the long-range force in $H_1$. This economical
description of the degrees of freedom will be quite useful when we consider the transport
equation at nonzero temperature. 

The remainder of the computations in this paper will be carried out using the Hamiltonian
$H$.
 
\section{Perturbation theory and Kubo formula}
\label{kubo}

We will begin in Section~\ref{sec:self}, 
as in the analysis for the bosonic case in I and in Ref~\onlinecite{s},
by an examination of the perturbative structure of the single particle self energy.
Then, we will proceed in Sections~\ref{sec:sxx} and~\ref{sec:sxy}
to an evaluation of $\widetilde{\sigma}_{xx}$ and $\widetilde{\sigma}_{xy}$
using the Kubo formula. We will gain an understanding of the physically different
contributions to $\widetilde{\sigma}$ by examining the structure of the bare
perturbation theory in powers of $\alpha$. The results for $\widetilde{\sigma}$
will have some unphysical
features which will finally be rectified by an analysis using the quantum Boltzmann
equation in the following Section~\ref{qtrans}.

\subsection{Self energy}
\label{sec:self}

The purpose of this section is to understand the structure of the mass 
renormalization in a single fermion propagator.
As noted earlier, we need to adjust the value of $M_0$ so that the renormalized
mass vanishes at $T=0$. However the renormalized mass need not vanish at $T>0$,
and we will describe its universal $T$ dependence here. We will denote the renormalized
mass by $M(T)$, and clearly we should have $M(0)=0$.

To leading order in $\alpha$,
it turns out to be sufficient to consider evaluation of $M(T)$ in a bare perturbation
theory in $\alpha$, without the need for any self-consistent mass renormalization;
the latter only modify results at higher orders in $\alpha$.
As $M_0$ is expected to be of order $\alpha$, we may therefore set $M_0=0$ within
the propagators that appear in the self energy.
To first order in $\alpha$, the fermion self energy for $H$ is
\begin{equation}
\Sigma ( \bk ) = 2 \pi \alpha i \epsilon_{ij} T \sum_{\epsilon_n}
\int \frac{d^2 p}{(2 \pi)^2} \frac{(k_i - p_i)}{(\bk - \bp)^2}
{\rm Tr} \left( \frac{1}{- i \epsilon_n + \bp \cdot \vec{\tau}} \tau_j
- \tau_j \frac{1}{- i \epsilon_n + \bp \cdot \vec{\tau}} \right).
\end{equation}
The trace is over the Dirac spinor space,
$\epsilon_n$ is a fermionic Matsubara frequency, and to this order,
$\Sigma$ is independent of the fermion frequency.
Evaluating the trace and performing the frequency summation, we get
\begin{equation}
\Sigma ( \bk ) = - \pi \alpha \tau_z \int \frac{d^2 p}{(2 \pi)^2}
\frac{(\bk - \bp)\cdot \bp}{p (\bk - \bp)^2} \tanh(p/2T)
\label{self1}
\end{equation}
The leading $\tau_z$ shows that $\Sigma (0)$ is precisely a mass renormalization,
and therefore
\begin{equation}
M(T) = M_0 + \left. \Sigma (0)\right|_{T=0}.
\label{self2}
\end{equation}
Using $M(0)=0$, we get
\begin{equation}
M_0 = - 2 \pi \alpha \int^{\Lambda} \frac{d^2 p}{(2 \pi)^2}
\frac{1}{p}
\label{self3}
\end{equation}
where $\Lambda$ is an ultraviolet momentum cutoff, whose nature
we will not have to precisely specify.
Finally, inserting (\ref{self3}) and (\ref{self1}) into (\ref{self2}) we get
\begin{equation}
M(T) = - 4\pi \alpha \int \frac{d^2 p}{(2 \pi)^2}
\frac{f^0 (p)}{p}
\label{self4}
\end{equation}
where $f^0 (p)$ is the Fermi function
\begin{equation}
f^0 (p) = \frac{1}{e^{p/T} + 1}.
\label{fermi}
\end{equation}
The integral in (\ref{self4}) is easily evaluated, and we get our final result
\begin{equation}
M(T) = - 2 \alpha T \ln (2).
\label{self5}
\end{equation}
We will find later that in some cases it is necessary to include this
renormalized mass in the fermion propagators, while in other cases we will
simply be able to use the zeroth order result $M(T) = 0$.
 
\subsection{$\widetilde{\sigma}_{xx}$}
\label{sec:sxx}

\begin{figure}
\epsfxsize=5.5in
\centerline{\epsffile{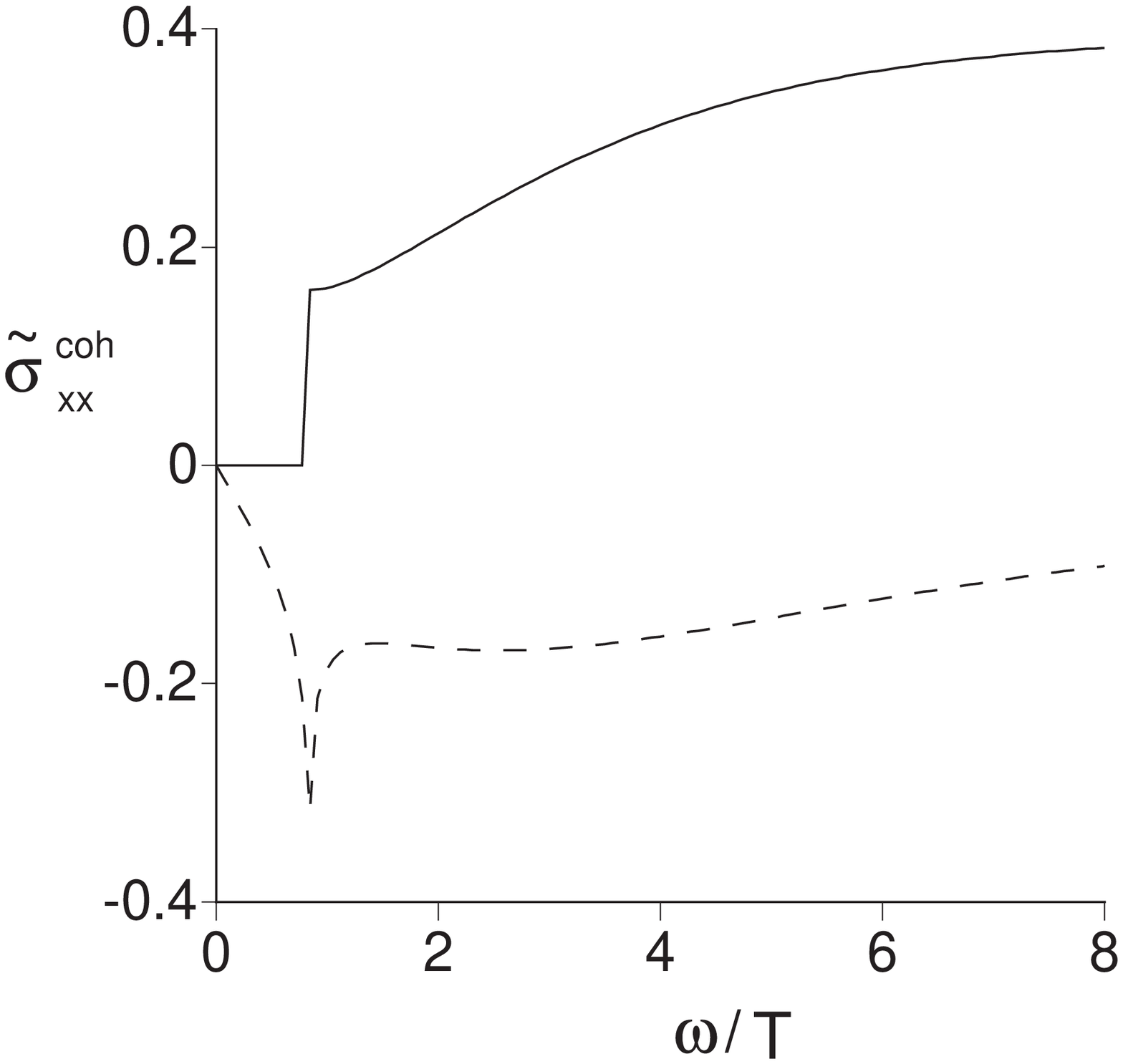}}
\vspace{0.2in}
\caption{Real (full line) and imaginary (dashed line)
parts of the perturbative result (\protect\ref{ecoh}) for the conductivity
$\widetilde{\sigma}_{xx}^{{\rm coh}} ( \omega )$ as a function of $\omega/T$
evaluated at $\alpha = 0.3$. The conductivity is measured in units
of $q^2 e^2/h$.
The singularity is at $\omega = 2 M (T)$, and is artifact of the absence
of damping at this order.
Note, as in (\protect\ref{sxxsum}), the
conductivity $\widetilde{\sigma}_{xx}$ contains another quasiparticle/quasihole
contribution which is a delta function in the perturbative approach.
}
\label{fig1}
\end{figure}

The leading contribution to $\widetilde{\sigma}_{xx}$ is given by
a single fermion
polarization bubble, with the mass $M(T)$ above included in the propagators.
To first order in $\alpha$ there are three additional diagrams, considered in Ref~\cite{cfw}, 
which contribute to the conductivity; however, these only yield contributions to
$\widetilde{\sigma}_{xy}$ and will be discussed in Section~\ref{sec:sxy}.
The fermion polarization contribution to $\widetilde{\sigma}_{xx}$ is
\begin{eqnarray}
\widetilde{\sigma}_{xx} ( i \omega_n ) = 
\frac{2 \pi q^2 e^2}{h \omega_n } T\sum_{\epsilon_n} \int \frac{d^2 k}{(2 \pi )^2}
 \mbox{Tr}&&
\left( 
\tau_x \frac{1}{(- i \epsilon_n + \bk \cdot \vec{\tau} -M(T) \tau_z)} \right. \nonumber\\
&&~~~~~~~~~~~~~\left. \times
\tau_x \frac{1}{(- i ( \epsilon_n + \omega_n) + 
\bk \cdot \vec{\tau}  - M(T) \tau_z )} \right).
\end{eqnarray}
The trace is over the
Dirac spinor space, and 
evaluating it and simplifying a bit, we obtain
\begin{equation}
\widetilde{\sigma}_{xx} ( i \omega_n ) = 
\frac{2 \pi q^2 e^2}{h \omega_n } T\sum_{\epsilon_n} 
\int \frac{d^2 k}{(2 \pi )^2}
\left[
\frac{4 k_x^2 + \omega_n^2}{(\varepsilon_k^2 + \epsilon_n^2)
(\varepsilon_k^2 + ( \epsilon_n + \omega_n)^2 )}
- \frac{2}{\varepsilon_k^2 + \epsilon_n^2} \right]
\label{sxx1}
\end{equation}
where
\begin{equation}
\varepsilon_k^2 \equiv k^2 + M^2 (T)
\end{equation}
The two terms in in the integrand in
(\ref{sxx1}) cancel at $T=0$ in the $\omega_n \rightarrow 0$ limit, but this is not
evident in the present form. To make this explicit, insert $1 = \partial k_x/\partial k_x$
in front of the second term, and integrate by parts under the $k_x$ integral;
this yields
\begin{equation}
\widetilde{\sigma}_{xx} ( i \omega_n ) = \frac{2 \pi q^2 e^2}{h \omega_n } 
T\sum_{\epsilon_n} \int \frac{d^2 k}{(2 \pi )^2}
\frac{1}{(\varepsilon_k^2 + \epsilon_n^2)}
\left[
\frac{4 k_x^2 + \omega_n^2}{(\varepsilon_k^2 + ( \epsilon_n + \omega_n)^2 )}
- \frac{4 k_x^2}{\varepsilon_k^2 + \epsilon_n^2} \right].
\end{equation}
We can now evaluate the summation over $\epsilon_n$ and analytically continue the result
to real $\omega$; as in I it is convenient to
separate the result into two distinct pieces
\begin{equation}
\widetilde{\sigma}_{xx} ( \omega ) \equiv
\widetilde{\sigma}_{xx}^{{\rm qp}} ( \omega ) + 
\widetilde{\sigma}_{xx}^{{\rm coh}} (\omega ).
\label{sxxsum}
\end{equation}
The first piece, $\widetilde{\sigma}^{{\rm qp}} ( \omega )$, is the contribution of 
thermally excited quasiparticles:
\begin{eqnarray}
\widetilde{\sigma}^{{\rm qp}}_{xx} ( \omega ) &=& 
\delta_{ij} \frac{4 \pi q^2 e^2}{h} \left[ \pi \delta ( \omega) + i {\cal P} \left( 
\frac{1}{\omega} \right)
\right] \int \frac{d^2 k}{(2 \pi)^2} \frac{k_x^2}{\varepsilon_k^2} \left( - \frac{\partial f^0 
(\varepsilon_k)}
{\partial \varepsilon_k} \right) \nonumber \\
 &=& 
\frac{q^2 e^2 {\cal N}(\alpha) }{h} \left[ \pi \delta ( \omega/T) + i {\cal P} \left( 
\frac{T}{\omega} \right) \right],
\label{eqp}
\end{eqnarray}
where ${\cal N}(\alpha)$, obtained after using (\ref{self5}), is a pure number
\begin{eqnarray}
{\cal N} ( \alpha ) &\equiv& \int_{2 \alpha \ln 2}^{\infty} d y
\frac{y^2 - 4 \alpha^2 \ln^2 (2)}{4 y \cosh^2 (y/2)} \nonumber \\
&=& \ln(2) + {\cal O}(\alpha^2)
\end{eqnarray}  
At this order, this contribution to the real part of the conductivity
is simply a delta function of frequency of weight $\propto ~T$ (which satisfies the scaling
form (\ref{scale})).
A more sophisticated quantum transport analysis (of the type carried out in I),
to be discussed in the next section, will show how this delta function is broadened
out by inelastic, incoherent collisions that appear at higher orders in $\alpha$.
The second piece, $\widetilde{\sigma}^{{\rm coh}} (\omega )$ is a smooth
continuum contribution from the creation of quasi-particle/quasi-hole
pairs by the external source, to the real part of the conductivity, and contains the 
coherent contribution in the limit $\omega/T \rightarrow \infty$;
after a change of variables in the momentum integration from $k$ to $\varepsilon_k$,
it can be written as
\begin{equation}
\widetilde{\sigma}^{{\rm coh}}_{xx} ( \omega ) =\frac{q^2 e^2}{2h}
\int_{|M(T)|}^{\infty} d \varepsilon_k 
\frac{-i \omega (\varepsilon_k^2 + M^2 (T))\tanh(\varepsilon_k /2T)}{\varepsilon_k 
( 4 \varepsilon_k^2 - 
(\omega+i\eta)^2)}, 
\label{ecoh}
\end{equation}
where $\eta$ is a positive infinitesimal.
The real part of this result can be evaluated in closed form
\begin{equation}
{\rm Re} \left[ \widetilde{\sigma}^{{\rm coh}}_{xx} ( \omega ) \right] = 
\frac{\pi q^2 e^2 }{8h }\theta ( |\omega| - 2 M(T))
 \left( 1 + \frac{4 M^2 (T)}{\omega^2} \right) \tanh(|\omega|/4T),
\end{equation}
while the imaginary part can be obtained by a Kramers-Kronig transform of 
${\rm Re} \left[ \widetilde{\sigma}^{{\rm coh}}_{xx} ( \omega ) \right] -
\pi q^2 e^2/8 h$.
Note that, after using (\ref{self5}) for $M(T)$, 
the result (\ref{ecoh}) is consistent with the
scaling form (\ref{scale}).
We show a plot of the frequency dependence of the real and imaginary
parts of $\widetilde{\sigma}_{xx}^{{\rm coh}}$ in Fig~\ref{fig1}. 
There are weak threshold singularities
at $\omega = 2M(T)$: a discontinuity in ${\rm Re}\left[\widetilde{\sigma}_{xx} \right]$
and a corresponding logarithmic divergence in ${\rm Im}\left[\widetilde{\sigma}_{xx} \right]$.
These are artifacts of the absence of damping in the fermion propagators at this order,
and the singularities are expected to be smoothed out at higher orders.

The previous computation of ${\rm Re}\left[ \widetilde{\sigma}_{xx} \right]$
in Ref~\onlinecite{cfw} was carried out at $T=0$,
in which we obtain only
the $\omega/T \rightarrow \infty$ limit of $\widetilde{\sigma}^{{\rm coh}}$
which, from (\ref{ecoh}), is given by
\begin{equation}
\widetilde{\sigma}_{xx} ( \omega /T \rightarrow \infty ) =  
\pi q^2 e^2/ 8 h.
\label{sxxlarge}
\end{equation}
In the limit $\omega /T \rightarrow 0$ relevant to d.c. transport, $\widetilde{\sigma}_{xx}$
is dominated by $\widetilde{\sigma}_{xx}^{\rm qp}$, which at this order is
a delta function at zero frequency: as stated earlier, this will be repaired in
Section~\ref{qtrans} by a quantum transport analysis. 

\subsection{$\widetilde{\sigma}_{xy}$}
\label{sec:sxy}

Unlike $\widetilde{\sigma}_{xx}$, the zeroth order in $\alpha$ result
for $\widetilde{\sigma}_{xy}$ vanishes identically at the critical point.
It is necessary to go to first order in $\alpha$, where
one finds two distinct types of contributions. The first 
(denoted $\widetilde{\sigma}_{xy}^{(1)}$)
comes from the
simple fermion polarization bubble considered above for $\widetilde{\sigma}_{xx}$,
but with the first order result for the mass $M(T)$ in (\ref{self5}) included.
The second comes from the diagrams, considered in Ref.~\onlinecite{cfw},
which are explicitly first order in $\alpha$: these account for the
momentum dependence of the self-energy (denoted $\widetilde{\sigma}_{xy}^{(2)}$)
and the corresponding vertex correction (denoted $\widetilde{\sigma}_{xy}^{(3)}$).

Considering first the contribution of the simple polarization bubble.
This gives
\begin{eqnarray}
\widetilde{\sigma}_{xy}^{(1)} ( i \omega_n ) = 
\frac{2 \pi q^2 e^2}{h \omega_n } T\sum_{\epsilon_n} \int \frac{d^2 k}{(2 \pi )^2}
 \mbox{Tr} &&
\left( 
\tau_x \frac{1}{(- i \epsilon_n + \bk \cdot \vec{\tau} - M(T) \tau_z) }
\right. \nonumber \\
&& ~~~~~~~~~~~~~~\times \left. \tau_y \frac{1}{(- i ( \epsilon_n + \omega_n) + 
\bk \cdot \vec{\tau}  - M(T) \tau_z )} \right).
\end{eqnarray}
Evaluating the Dirac trace and the frequency summation, we find that the 
leading $1/\omega_n$ explicitly cancels out: as a result there will be no
singular $\delta (\omega)$ contributions to $\widetilde{\sigma}_{xy}$ as
there were to $\widetilde{\sigma}_{xx}$. We change variables in the momentum 
integration from $k$ to $\varepsilon_k$, and analytically continue to real frequencies,
and the answer takes the final form
\begin{equation}
\widetilde{\sigma}_{xy}^{(1)} ( \omega ) = 
- \frac{2 q^2 e^2 M(T)}{h} \int_{|M(T)|}^{\infty}
d \varepsilon_k \frac{\tanh(\varepsilon_k /2T)}{4 \varepsilon_k ^2 - (\omega+ i \eta)^2}
\label{sxy1}
\end{equation}
Notice that there is $M(T)$ in the prefactor, and so this result is at least
first order in $\alpha$. Nevertheless, it is not permissible to simply set
$M(T) = 0$ in the lower limit of the integration to obtain the leading result: this is
because the integrand has a potential logarithmic divergence for $\omega_n =0$,
which is cutoff by the lower limit.
The result (\ref{sxy1}) vanishes as 
$\sim i T/\omega $ in the limit $\omega/T \rightarrow \infty$, 
while in the d.c. limit $\omega/T \rightarrow$ we get after using (\ref{self5})
and manipulations described in Appendix~\ref{zeta}:
\begin{eqnarray}
\widetilde{\sigma}_{xy}^{(1)} ( \omega/T \rightarrow 0)
= && \frac{q^2 e^2}{h} \left[ \left(\frac{\ln(2)}{2}\right) \alpha \ln (1/\alpha)  
 \right. \nonumber \\
&&~+ \left. \frac{\ln(2)}{6} \left\{
3 - 3 \gamma - 7 \ln(2) - 3 \ln(\ln(2)) -36 \zeta^{\prime} (-1)
\right\} \alpha +
{\cal O} ( \alpha^3 )
\right]
\label{sxy1n}
\end{eqnarray}
Here $\gamma = 0.5772\ldots$ is Euler's constant and $\zeta (s)$ is the
Reimann zeta function.
Notice the non-analytic $\alpha \ln ( 1/\alpha)$ term: the proper $T$ dependence of the mass
$M(T)$ discussed in Section~\ref{sec:self} was required to obtain this.

We now consider the
three diagrams for  $\widetilde{\sigma}$ which are explicitly first order in
$\alpha$. These were considered in Ref~\onlinecite{cfw} at $T=0$;
the $T>0$ calculations are
forbidingly complicated in the Lorentz gauge used in Ref~\onlinecite{cfw} because
of the retardation in the gauge field propagator; they simplify considerably in the
Coulomb gauge we are using here. 
To first order in $\alpha$ we can set $M(T)=0$ in the propagators of these diagrams;
this is justified {\em a posteriori} by the absence of infrared divergences in
our evaluation of these diagrams.
The first such 
contribution is the sum of the two diagrams accounting for the momentum
dependence of the self-energy
correction 
in one of the fermion propagators: 
\begin{eqnarray}
\widetilde{\sigma}_{xy}^{(2)} ( i \omega_n ) = &&
\frac{8 \pi^2 q^2 e^2 \alpha}{h \omega_n} i \epsilon_{ij} 
T \sum_{\epsilon_n} \int \frac{d^2 k}{(2 \pi)^2} \frac{d^2 p}{(2 \pi)^2}
\left[\frac{(k_i - p_i)}{(\bk - \bp)^2} + \frac{p_i}{p^2} \right] \nonumber \\
&&~~~~
\times {\rm Tr} \left[
\tau_x \frac{1}{(-i \epsilon_n + \bk \cdot \vec{\tau})}
\left\{ \frac{1}{(-i  \Omega_n + \bp \cdot \vec{\tau})} \tau_j
- \tau_j \frac{1}{(-i \Omega_n + \bp \cdot \vec{\tau})}\right\}
\right.
\nonumber \\
&&~~~~~~~~~~~~~~~~~~~~~\left.
\frac{1}{(-i \epsilon_n + \bk \cdot \vec{\tau})} \tau_y  \frac{1}{(-i 
(\epsilon_n + \omega_n) + \bk \cdot \vec{\tau})} 
 \right] 
\label{self}
\end{eqnarray}
The second term in the first square bracket is the subtraction that corrects
for the momentum-independent mass renormalization that was already accounted for
by including a $M(T)$ in $\widetilde{\sigma}^{(1)}_{xy}$.
The final contribution to $\widetilde{\sigma}_{xy}$ is the first-order vertex correction
corresponding to the self-energy in $\widetilde{\sigma}_{xy}^{(2)}$:
\begin{eqnarray}
\widetilde{\sigma}_{xy}^{(3)} ( i \omega_n ) = &&
 \frac{4 \pi^2 q^2 e^2 \alpha}{h \omega_n} i \epsilon_{ij} 
T^2 \sum_{\epsilon_n, \Omega_n} \int \frac{d^2 k}{(2 \pi)^2} \frac{d^2 p}{(2 \pi)^2}
\frac{(k_i - p_i)}{(\bk - \bp)^2} \nonumber \\
&&~~\times {\rm Tr} \left[
\tau_x \frac{1}{(-i \epsilon_n + \bk \cdot \vec{\tau})}
\left\{
 \frac{1}{(-i \Omega_n + \bp \cdot \vec{\tau})} \tau_y
\frac{1}{(-i (\Omega_n + \omega_n) + \bp \cdot \vec{\tau})} \tau_j \right.\right.
\nonumber \\
&&~~~~~\left. \left.-
\tau_j \frac{1}{(-i \Omega_n + \bp \cdot \vec{\tau})} \tau_y
\frac{1}{(-i (\Omega_n + \omega_n) + \bp \cdot \vec{\tau})} \right\}
\frac{1}{(-i (\epsilon_n + \omega_n) + \bk \cdot \vec{\tau})}
 \right] 
\label{vertex}
\end{eqnarray}
We now evaluate the trace over Dirac indices, sum over 
frequencies in (\ref{self}) and analytically continue to real frequencies to obtain:
\begin{eqnarray}
\widetilde{\sigma}_{xy}^{(2)} (\omega ) &=& 
-\frac{8 \pi^2 q^2 e^2 \alpha}{h}  \int \frac{d^2 k}{(2 \pi)^2} \frac{d^2 p}{(2 \pi)^2}
\frac{\tanh(p/2 T) \tanh(k/2 T)}{pk( 4 k^2 - (\omega+i\eta)^2)}
\left(
\frac{(p^2 - \bk\cdot \bp)}{(\bk - \bp)^2} - 1 \right) \nonumber \\
&=& \frac{2 q^2 e^2 \alpha}{h} \int_0^{\infty} dp
\int_{p}^{\infty} dk 
\frac{\tanh(p/2 T) \tanh(k/2 T)}{( 4 k^2 - (\omega + i \eta )^2 )},
\label{sxy2}
\end{eqnarray}
where in the second expression we have evaluated the angular integrals.
As discussed in Ref~\cite{cfw}, it is important in (\ref{sxy2})
that the $k$ integral be evaluated first as it is convergent. There is a logarithmic
divergence in the subsequent $p$ integral but this will cancel against an opposite
contribution in $\widetilde{\sigma}_{xy}^{(3)}$.
The trace and summation over frequencies in (\ref{vertex}) yields
\begin{eqnarray}
\widetilde{\sigma}_{xy}^{(3)} ( \omega ) &=& 
-\frac{32 \pi^2 q^2 e^2 \alpha}{h} 
\int \frac{d^2 k}{(2 \pi)^2} \frac{d^2 p}{(2 \pi)^2}
\frac{(\bk \times \bp)^2 }{p k (\bk - \bp)^2}
\frac{\tanh(p/2T) \tanh(k/2 T)}{(4 k^2 - (\omega+i\eta)^2 ) ( 4 p^2 - (\omega+ i \eta)^2)}
\nonumber \\
&=& - \frac{4 q^2 e^2 \alpha}{h}
\int_0^{\infty} dp
\int_{0}^{\infty} dk~ 
{\rm Min}(k^2, p^2)
\frac{\tanh(p/2T) \tanh(k/2 T)}{( 4 k^2 - (\omega + i \eta)^2 ) ( 4 p^2
- (\omega + i \eta)^2 )}
\label{sxy3}
\end{eqnarray}
where again the second expression is obtained after evaluating the angular integrals,
and we choose to evaluate the $k$ integral first.
The integrals can be evaluated explicitly in the limit $\omega /T \rightarrow \infty$,
and we get 
\begin{equation}
\widetilde{\sigma}_{xy}^{(2)}( \omega/T \rightarrow \infty)
 + \widetilde{\sigma}_{xy}^{(3)}( \omega/T \rightarrow \infty) 
 = \frac{q^2 e^2 \alpha}{h} 
  \frac{(\pi^2 +4 )}{16}
\end{equation}
In the opposite limit, $\omega /T \rightarrow 0$, we evaluated the integrals numerically, and 
found to a very high accuracy that
\begin{equation}
\widetilde{\sigma}_{xy}^{(2)}( \omega/T \rightarrow 0)
 + \widetilde{\sigma}_{xy}^{(3)}( \omega/T \rightarrow 0) 
  = \frac{q^2 e^2 \alpha}{4h},
\end{equation}
although we have no analytic proof of this simple result.
The full frequency dependence is most easily evaluated by first taking the
imaginary parts of (\ref{self}) and (\ref{vertex}), which can be shown to
simplify to
\begin{equation}
{\rm Im} \left[ \widetilde{\sigma}_{xy}^{(2)} ( \omega ) + 
\widetilde{\sigma}_{xy}^{(3)} ( \omega ) \right] = \frac{q^2 e^2 \pi \alpha}{h}
\omega \tanh( |\omega|/4 T) \int_0^{\infty} dp f^0 ( p/T) {\cal P}
\left( \frac{1}{4 p^2 - \omega^2} \right).
\label{imsxy2}
\end{equation}
The real part is obtained by adding the constant $q^2 e^2 \alpha ( \pi^2 + 4)/16 h$
to the Kramers-Kronig transform of (\ref{imsxy2}).

We now combine (\ref{sxy1}), (\ref{sxy2}) and (\ref{sxy3})
to obtain the final results for $\widetilde{\sigma}_{xy}$.
We have the limiting values
\begin{eqnarray}
\widetilde{\sigma}_{xy} ( \omega/T \rightarrow 0)
&=& \frac{q^2 e^2}{h} \left[ \left(\frac{\ln(2)}{2}\right) \alpha \ln (1/\alpha)  + 
(0.650988\ldots)\alpha +
{\cal O} ( \alpha^3 )
\right] \nonumber \\
\widetilde{\sigma}_{xy} ( \omega/T \rightarrow \infty)
&=& \frac{q^2 e^2}{h} \left[ \left( \frac{\pi^2 +4 }{16}
\right) \alpha +
{\cal O} ( \alpha^3 )
\right]
\label{sxylim}
\end{eqnarray}
The second of these results has been obtained previously~\cite{cfw,st}; that
calculation was done in a Lorentz invariant gauge, and it is reassuring that
the same result has been obtained here in a completely different, non-relativistic
gauge.
Notice that there are no delta function contributions to $\widetilde{\sigma}_{xy}$:
this implies that the entire contribution is due to coherent quasi-particle/quasi-hole
creation, and the collisionless flow of pre-existing 
thermally excited particles/holes contributes only to $\widetilde{\sigma}_{xx}^{{\rm qp}}$

\begin{figure}
\epsfxsize=5.5in
\centerline{\epsffile{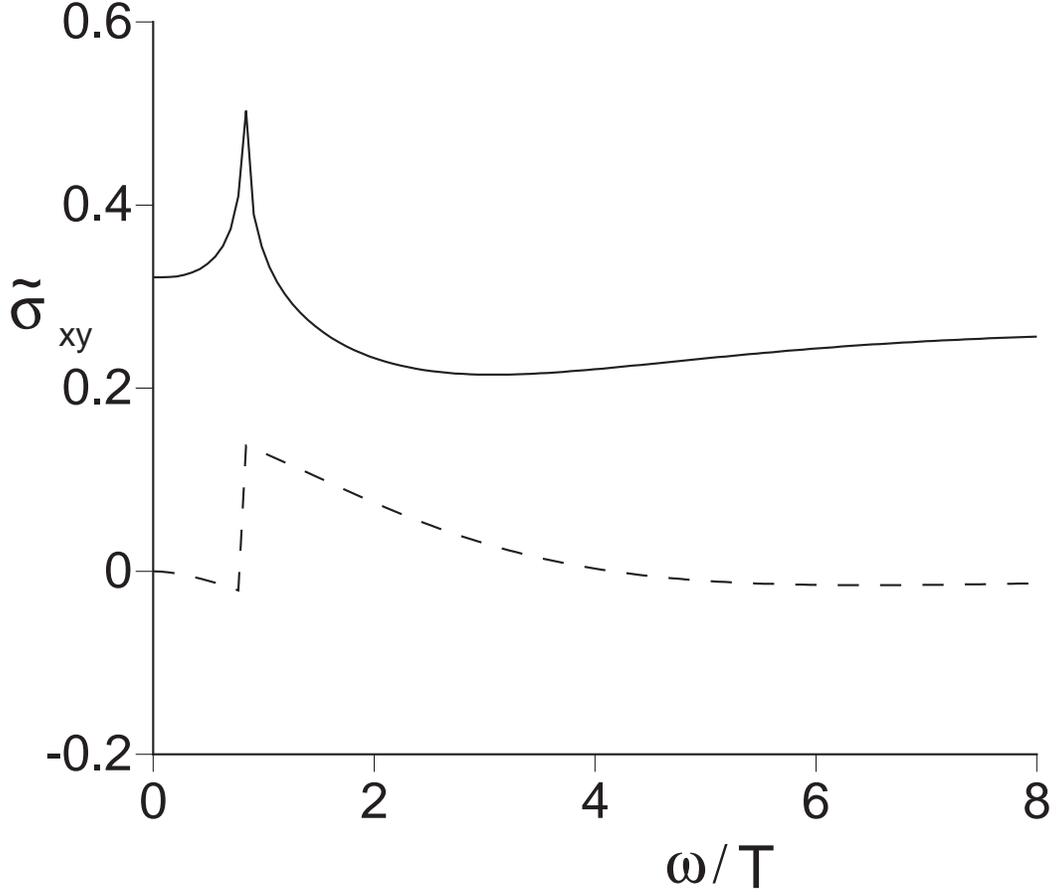}}
\vspace{0.2in}
\caption{Real and imaginary parts of the perturbative results for the Hall 
conductivity (\protect\ref{sxy1}),
(\protect\ref{sxy2}) and (\protect\ref{sxy3}) for 
$\widetilde{\sigma}_{xy} ( \omega ) = 
\widetilde{\sigma}_{xy}^{(1)} + \widetilde{\sigma}_{xy}^{(2)} 
+\widetilde{\sigma}_{xy}^{(3)}$ as a function of the $\omega/T$
evaluated at $\alpha = 0.3$. The conductivity is measured in units
of $q^2 e^2/h$.
Again the spurious singularities are at $\omega = 2 M(T)$.
Note that unlike, $\widetilde{\sigma}_{xx}$, the Hall
conductivity is adequately described by the perturbation theory.
}
\label{fig2}
\end{figure}

We conclude this section be showing
a plot of the frequency dependence of the real and imaginary
parts of $\widetilde{\sigma}_{xy}$ in Fig~\ref{fig2}. 
Notice that, as in $\widetilde{\sigma}_{xx}$ there are threshold singularities
at $\omega = 2M(T)$: a discontinuity in ${\rm Im}\left[\widetilde{\sigma}_{xx} \right]$
and a corresponding logarithmic divergence in ${\rm Im}\left[\widetilde{\sigma}_{xx} \right]$.
Again these artifacts of the absence of damping in the fermion propagators at this order.

\section{Quantum Transport Equations}
\label{qtrans}

The perturbative analysis of $\widetilde{\sigma}$ in Section~\ref{kubo}
produced results which had singularities at two frequencies: a delta function
in $\widetilde{\sigma}_{xx}^{{\rm qp}}$ at $\omega =0$, and much weaker
logarithmic singularities at $\omega = 2 M(T)$. Both these features are in fact 
artifacts of the low order results, and disappear when damping of
fermion excitations is accounted for at higher orders. 
In this section, we shall show how the delta function singularity
at $\omega = 0$ is broadened out at order $\alpha^2$. 
The singularity at $\omega = 2 M(T)$ is due to a threshold to creation of 
particle-hole pairs: it is also expected to be smoothed out
at order $\alpha^2$, but determination of the precise form of this is left as
an open problem.

So to reiterate, this section will only describe the changes to
the result for 
$\widetilde{\sigma}_{xx}^{{\rm qp}}$  in (\ref{eqp})
at higher order in $\alpha$. The perturbative results for the other
pieces of $\widetilde{\sigma}$, in particular,
(\ref{ecoh}) for $\widetilde{\sigma}_{xx}$ and (\ref{sxy1}), (\ref{sxy2}),
and (\ref{sxy3}) for $\widetilde{\sigma}_{xy}$, will be used unchanged
in our subsequent considerations as their spurious singularities
are relatively minor.

The delta function singularity in
$\widetilde{\sigma}_{xx}^{{\rm qp}}$, will be explicitly shown below
to be due to the collisionless transport of thermally excited quasiparticles and
quasiholes. This will be done as in I: by writing down transport equations for
the distribution functions of these excitations. The collisionless form of these
equations appear in Section~\ref{cless} and their solution reproduces the
result for $\widetilde{\sigma}_{xx}^{{\rm qp}}$ obtained in Section~\ref{sec:sxx};
then, in Section~\ref{cdom}, we include 
collision terms and present the numerical solution.

\subsection{Collisionless transport of quasiparticles and quasiholes}
\label{cless}

We begin by recalling the defining relation (\ref{cs9}) for $\widetilde{\sigma}$:
it is the linear response in the current to the total internal field ${\bf E}_T$.
Fortunately, the field ${\bf E}_T$ appears in the transport
equation in a simple and natural way: the effect of the Hartree and Fock terms
is precisely such as to modify the external field ${\bf E}$ to ${\bf E}_T$~\cite{kb}.
As a result we can simply omit these terms while replacing ${\bf E}$ by
${\bf E}_T$, and 
determination of the $\widetilde{\sigma}$ becomes, as in I, a matter
of balancing the ballistic, particle streaming terms with the collision term.

Throughout Section~\ref{qtrans} we will work with fermionic excitations with
a spectrum $\varepsilon_k = k$ {\em i.e.} we will set the mass $M(T)=0$;
as the Hartree and Fock terms
in the self-energy are also implicitly accounted for in our analysis, this is
equivalent to simply disregarding such terms and working with a $H_O$ with $M_0 = 0$.
In this approximation the
transport analysis will yield a
normalization constant for the spectral weight in the conductivity
in (\ref{eqp}) which is given by the result correct to leading order in $\alpha$,
${\cal N} (\alpha) = \ln(2)$. 
Our objective will then be to describe the broadening of the delta function
in (\ref{eqp}), while omiting all terms which lead to corrections in
its total spectral weight.

The simplest formulation of the transport equations is in a basis which
diagonalizes the Hamiltonian $H_0$. To do this, we first 
express $\psi$ in its Fourier components
\begin{equation}
\psi ( \bx, t) = \int \frac{d^2 k}{(2 \pi )^2} \left(
\begin{array}{c}
c_1 ( \bk,t) \\
c_2 ( \bk,t) \end{array}
\right) e^{i \bk \cdot \bx} ,
\end{equation}
and then perform a unitary transformation from the Fourier mode operators
$(c_1 , c_2)$ to $(\gamma_+ , \gamma_-)$:
\begin{eqnarray}
c_1 (k) &=& \frac{1}{\sqrt{2}} ( \gamma_+ (\bk) + \gamma_- (\bk)) \nonumber \\
c_2 (k) &=& \frac{K}{\sqrt{2} k} ( \gamma_+ (\bk) - \gamma_- (\bk)).
\end{eqnarray}
We have introduced here a notational convention that we shall find quite useful in
the following: as $\bk$ is a two-dimensional momentum, we can define the complex 
number $K$ by
\begin{equation}
K \equiv k_x + i k_y~~~~\mbox{where}~~~~~\bk \equiv (k_x, k_y)
\end{equation}
and $k = |\bk| = |K|$.
Expressing
the Hamiltonian $H_0$ in terms of $\gamma_{\pm}$, we obtain the simple result
\begin{equation}
H_0 = \sum_{\lambda} \int \frac{d^2 k}{(2 \pi )^2} 
 \lambda k \gamma_{\lambda}^{\dagger} (\bk) \gamma_{\lambda} ( \bk )
\end{equation}
where the sum over $\lambda$ extends over $+,-$.
So $\gamma_+^{\dagger}$ creates a particle with energy $k$, while 
$\gamma_-^{\dagger}$ creates a particle with energy $-k$. At zero temperature
the negative energy states will be filled: so we identify $\gamma_+^{\dagger}$
as the creation operator for quasiparticles, while $\gamma_-$ is the
creation operator for quasiholes. Notice that the quasiparticles and
quasiholes are distinct excitations in different bands, and the absence of a quasiparticle
is {\em not} equivalent to the presence of a quasihole: this latter feature was also 
present in the bosonic analysis of I, and as argued there, allows the system to acquire
a non-zero d.c. resistance while maintaining conservation of total momentum.

Let us also, for future use, express the interaction Hamiltonian $H_1$ in terms
of the $\gamma_{\pm}$:
\begin{eqnarray}
&& H_1 =  \sum_{\lambda_1 \lambda_2 \lambda_3 \lambda_4}
\int \frac{d^2 k_1 }{(2 \pi )^2}
\frac{d^2 k_2 }{(2 \pi )^2}
\frac{d^2 k_q }{(2 \pi )^2} \nonumber \\
&&~~~~~~~~~~~~~~~~~~~~~~~~~~~
\times T_{\lambda_1 \lambda_2 \lambda_3 \lambda_4} (\bk_1 , \bk_2 , \bq )
\gamma_{\lambda_4}^{\dagger} ( \bk_1+\bq ) 
\gamma_{\lambda_3}^{\dagger} ( \bk_2-\bq )
\gamma_{\lambda_2} ( \bk_2 )
\gamma_{\lambda_1} ( \bk_1 )
\end{eqnarray}
where
\begin{equation}
T_{\lambda_1 \lambda_2 \lambda_3 \lambda_4} (\bk_1 , \bk_2 , \bq)
= - \frac{\pi \alpha}{2 q^2}
\left[ 1 + \lambda_1 \lambda_4 \frac{(K_1^{\ast} + Q^{\ast}) K_1}{|\bk_1
+ \bq| k_1} \right] 
\left[
\lambda_3 \frac{Q (K_2^{\ast} - Q^{\ast})}{|\bk_2 - \bq|} -
\lambda_2 \frac{Q^{\ast} K_2}{k_2} \right]
\label{deft}
\end{equation}
Notice that the scattering amplitude, $T$, is singular for forward scattering:
$T \sim 1/q$. This is a characteristic property of scattering between anyons~\cite{wilczek}.
We will have to keep careful track of this potential infrared
singularity in our subsequent transport analysis.

Finally, we also express the electrical current, defined in (\ref{cs6}),
in terms of the $\gamma_{\pm}$. For the case of a spatially independent
current (which is the only case of interest here), the result can be written as
\begin{equation}
{\bf J} = {\bf J}_I + {\bf J}_{II}
\end{equation}
with
\begin{equation}
{\bf J}_I =  qe\sum_{\lambda} \int \frac{d^2 k}{(2 \pi )^2}
\frac{\lambda \bk}{k} \gamma_{\lambda}^{\dagger} (\bk) \gamma_{\lambda} ( \bk )
\label{defj1}
\end{equation}
and
\begin{equation}
{\bf J}_{II} = - iqe \int \frac{d^2 k}{(2 \pi )^2}
\frac{(\hat{{\bf z}} \times \bk)}{k} \left(
\gamma_{+}^{\dagger} (\bk) \gamma_{-} ( \bk ) - 
\gamma_{-}^{\dagger} (\bk) \gamma_{+} ( \bk
)
\right)
\end{equation}
where $\hat{{\bf z}}$, a unit vector orthogonal to the $x,y$ plane.
As in I, ${\bf J}_I$ measures the current carried by motion of the quasiparticles
and quasiholes---notice the $\lambda$ prefactor, indicating that these excitations
have opposite charges. The operator ${\bf J}_II$ creates a quasiparticle-quasihole
pair: throughout this section we will work in the approximation in which ${\bf J}_{II} = 0$.
We will also neglect all mixing between the $\left\langle
\gamma_{\pm}^{\dagger} \gamma_{\pm} \right \rangle$ and 
$\left\langle
\gamma_{\pm}^{\dagger} \gamma_{\mp} \right \rangle$ distribution functions.
These approximations simply amount to neglecting all but the 
$\widetilde{\sigma}_{xx}^{{\rm qp}}$ component of $\widetilde{\sigma}$---the other
components were already computed satisfactorily in perturbation theory in Section~\ref{kubo}.

We can now use the standard equation of motion analysis~\cite{kb} to write
down the collisionless transport equations for the excitations. As a first step,
we define the distribution functions
\begin{equation}
f_{\lambda} ( \bk , t)  = \left\langle \gamma_{\lambda}^{\dagger}
( \bk, t) \gamma_{\lambda} ( \bk , t) \right \rangle.
\label{defg}
\end{equation}
In equilibrium in the absence of external perturbations, these are related to
the Fermi function defined in (\ref{fermi}) 
\begin{eqnarray}
f_{+} ( \bk, t ) &=& f^{0} ( k )  \nonumber \\
f_{-} ( \bk, t ) &=& f^{0} ( - k ) =  1 - f^{0} (k)
\end{eqnarray}
Then to first order in $\alpha$, in the presence of an external electric
field ${\bf E}$, we find the simple equations
\begin{equation}
\left( \frac{\partial}{\partial t} +  q e {\bf E}_T \cdot \frac{\partial}{\partial
\bk} \right)
 f_{\lambda} ( \bk, t) = 0.
\label{trans0}
\end{equation}
At this order, in the approximations discussed above, the only effect of the
interactions has been to replace ${\bf E}$ by ${\bf E}_T$.
It is a simple matter to solve (\ref{trans0}) in linear response. First we parametrize
the change in $f_{\lambda}$ from its equilibrium value by
\begin{equation}
f_{\lambda} ( \bk , \omega ) = 2 \pi \delta(\omega) f^{0} ( \lambda k )
 + \lambda q e \bk \cdot {\bf E}_T ( \omega ) g (k, \omega ),
\label{paramet}
\end{equation}
where we have performed a Fourier transform in time to frequencies, $\omega$,
and introduced the unknown function $g(k, \omega)$.
Notice that the change in the distribution functions has an opposite sign
for quasiparticles and quasiholes. Thus the quasiparticles and quasiholes
move in opposite directions: as they have opposite charges, their electrical
currents are equal, while their net momenta have opposite signs.
Inserting (\ref{paramet}) into (\ref{trans0}), we can obtain a simple
solution for the function $g$
\begin{equation}
g(k,\omega ) = \frac{1}{(- i \omega + \eta) k} \left(- 
\frac{\partial f^0 (k)}{\partial k} \right).
\end{equation}
Inserting this result into (\ref{defg}) and (\ref{defj1}), we obtain
the conductivity
\begin{equation}
\widetilde{\sigma}_{xx}^{{\rm qp}} ( \omega ) = 
2 \frac{2 \pi q^2 e^2}{(-i \omega+\eta) h} \int \frac{d^2 k}{(2 \pi)^2} 
\frac{k_x^2}{k^2} \left( - \frac{\partial f^0 (k)}{\partial k} \right),
\label{eqp1}
\end{equation}
which is precisely our earlier result (\ref{eqp}), in the approximation
$M(T) =0$ and ${\cal N} ( \alpha ) = \ln(2)$.
In this approach, it is now evident that the leading factor of 2 comes
from the sum over $\lambda$: {\em i.e.} the quasiparticles and quasiholes
contribute equally to the total current.

\subsection{Collision-dominated transport}
\label{cdom}

We now include collision terms on the right hand side of (\ref{trans0}).
As in I, while the terms can be formally derived using the methods of Ref~\onlinecite{kb},
it is easier to determine by a simple application of Fermi's golden rule. In the latter
manner we obtained
\begin{eqnarray}
&& \left( \frac{\partial}{\partial t} +  qe {\bf E}_T \cdot \frac{\partial}{\partial
\bk} \right)
 f_{\lambda} ( \bk, t) = \nonumber \\
&&~~~~~~~-  \int \frac{d^2 k_1}{(2 \pi)^2} 
\frac{d^2 q}{(2 \pi)^2} 
 (2 \pi) \delta (k + k_1 - |\bk+\bq| - |\bk_1 - \bq|)
\Biggl\{ \nonumber \\
&&~~~~~~~~~~~~~ \left| T_1
\right|^2 \Bigl\{  f_{\lambda} ( \bk , t) f_{-\lambda} (-\bk_1 + \bq , t) 
[ 1 - f_{\lambda} ( \bk+\bq , t)] [ 1 - f_{-\lambda} ( -\bk_1 , t)] \nonumber \\
&&~~~~~~~~~~~~~~~~~~~~~~~~~~~~~~~-  
[1 - f_{\lambda} ( \bk , t)][1 - f_{-\lambda} (-\bk_1+\bq ,
t)] f_{\lambda} ( \bk+\bq , t) f_{-\lambda} (-\bk_1 , t) \Bigr\} \nonumber \\
&&~~~~~~~~~~~~~ + 
\frac{1}{2} \left| T_{2}
\right|^2  \Bigl\{ f_{\lambda} ( \bk , t) f_{\lambda} (\bk_1 , t)  [ 1 - f_{\lambda} ( \bk+\bq
, t)][ 1 - f_{\lambda} ( \bk_1-\bq , t)] \nonumber \\ 
&&~~~~~~~~~~~~~~~~~~~~~~~~~~~~~~~- [1 - f_{\lambda} ( \bk , t)][1 -
f_{\lambda} (\bk_1 , t)] f_{\lambda} ( \bk+\bq , t) f_{\lambda} ( \bk_1-\bq , t) \Bigr\}
\Biggr\}.
\label{trans1}
\end{eqnarray}
where
\begin{eqnarray}
T_1  \equiv && T_{+--+}( \bk, -\bk_1 + \bq, \bq)
- T_{-+-+} ( -\bk_1 + \bq, \bk, \bk + \bk_1) \nonumber \\
&&~~~~~~~-
T_{+-+-} ( \bk, -\bk_1 + \bq, -\bk-\bk_1 )
+ T_{-++-} ( -\bk_1 + \bq, \bk, -\bq ) 
\nonumber \\
T_2 ( \bk, \bk_1, \bq) \equiv && T_{++++}( \bk, \bk_1, \bq)
- T_{++++} ( \bk_1 , \bk, \bk +\bq - \bk_1) \nonumber \\
&&~~~~~~~-
T_{++++} ( \bk, \bk_1 ,\bk_1 -\bk-\bq )
+ T_{++++} ( \bk_1 , \bk, -\bq ) 
\label{deft12}
\end{eqnarray}
The terms proportional to $|T_1|^2$ represent collisions between oppositely charged
particles, while those proportional to $|T_2|^2$ are collisions between like
charges. There are also processes a particle-hole pair is created: as in I, these can
be dropped because they have vanishing
phase space upon imposition of the energy conservation constraint with dispersion
$\varepsilon_k = k$

We now proceed to the linearization of (\ref{trans1}) by inserting the parametrization
(\ref{paramet}) and find
\begin{eqnarray}
&& -i \omega k g (k, \omega ) + \frac{\partial f^0 (k)}{\partial k}
= \nonumber \\
&&~~~~~~~-  \int \frac{d^2 k_1}{(2 \pi)^2} 
\frac{d^2 q}{(2 \pi)^2} 
 (2 \pi) \delta (k + k_1 - |\bk+\bq| - |\bk_1 - \bq|)
\Biggl\{ \nonumber \\
&&~~~~~~~~~~~~~~~~   \left(\frac{(e^{(|\bk_1 - \bq| + |\bk +\bq|)/T}+ e^{ k_1 /T})
(|T_1|^2 + |T_2|^2 /2)}{(
e^{k_1 /T} + 1)(e^{|\bk_1 - \bq|/T}+1)(e^{|\bk + \bq|/T}+1)} \right) k g(k, \omega) \nonumber \\
&&~~~~~~~~~~~~~~~- \left(\frac{(e^{(|\bk_1 - \bq|+ |\bk +\bq|)/T} + e^{k/T})
(|T_1|^2 - |T_2|^2 /2)}{(
e^{k/T} + 1)(e^{|\bk_1 - \bq|/T}+1)(e^{|\bk + \bq|/T}+1)} \right)
\left( \frac{\bk\cdot \bk_1}{k}\right)
g(k_1, \omega) \nonumber \\
&&~~~~~~~~~~~~~~~+ \left(\frac{(e^{|\bk +\bq|/T} + e^{(k+k_1)/T})
(|T_1|^2 - |T_2|^2 /2)}{(
e^{k/T} + 1)(e^{k_1/T}+1)(e^{|\bk + \bq|/T}+1)}\right)
\left( \frac{\bk\cdot (\bk_1-\bq)}{k}\right)
g(|\bk_1-\bq|, \omega) \nonumber \\
&&~~~~~~~~~~~~~~~- \left(\frac{(e^{|\bk_1 - \bq|/T} + e^{(k+k_1)/T})
(|T_1|^2 + |T_2|^2 /2)}{(
e^{k/T} + 1)(e^{k_1 /T}+1)(e^{|\bk_1 - \bq|/T}+1)}\right)
\left( \frac{\bk\cdot (\bk+\bq)}{k}\right)
g(|\bk+ \bq|, \omega) \Biggr\}.
\label{trans2}
\end{eqnarray}
The final ingredient necessary before assembling all the results of this
paper is now a solution of (\ref{trans2}) for the distribution function $g(k, \omega)$.
An analytic solution is clearly out of the question and, we obtained the solution
numerically. However, before attempting any numerical solution, the singular nature
of the scattering cross-sections $|T_1|^2$ and $|T_2|^2$ makes it essential
to obtain an analytic understanding of the nature of any potential infrared divergences
arising from the momentum integration. We will show below that all the potentially
infrared divergent
terms in the collision integral do indeed vanish. However just 
this knowledge is not sufficient to allow one to proceed with the numerical integration.
One does not want numerically integrate an integrand with potentially
divergent terms which
will eventually average out to a finite answer, 
as such a procedure is clearly prone to large systematic numerical errors;
rather one wants an integrand which is explicitly free of singularities.
On the other hand, the integrand is dependent upon the unknown function $g$,
so explicitly removing the singularities is difficult. 
These issues are clearly crucial, but also vexing, and stymied the author for some time.
Eventually, a remarkably simple and useful parametrization of the collision term
was found. As the analytic discussion
of the cancellation of singularities is also simplest in this new parametrization,
we will first introduce it, and then discuss the cancellation.

The key is to carry out the $\bq$ integration in the collision term
of (\ref{trans2}) first, and to 
notice that the delta function associated with energy conservation
\begin{equation}
\delta (k + k_1 - |\bk+\bq| - |\bk_1 - \bq|)
\label{defellipse}
\end{equation}
defines an ellipse in the $\bq$ plane with foci at $-\bk$ and $\bk_1$. 
It therefore pays to parametrize $\bq$
in an {\em elliptic\/} co-ordinate system~\cite{mf}.
We introduce elliptic co-ordinates
\begin{equation}
0 \leq \mu < \infty~~~~~,~~~~~-\pi < \theta \leq \pi
\end{equation}
and express $\bq$ in terms of $\mu$ and $\theta$ by
\begin{equation}
Q = \frac{K_1 - K}{2} + \frac{(K_1 + K)}{4} \left(
e^{\mu+i\theta} + e^{-\mu-i\theta} \right).
\label{Qval}
\end{equation}
Here $\bk$ and $\bk_1$ are regarded as fixed external parameters, and $\bk_1$ will
be subsequently integrated over.
Then it can be easily verified that the ellipse represented by (\ref{defellipse})
is given simply by
\begin{equation}
\cosh \mu = \frac{k + k_1}{|\bk + \bk_1|}
\label{muval}
\end{equation}
The integral over $\bq$ in (\ref{trans2}) is a two-dimensional integral over
$\mu$ and $\theta$, and the integral over $\mu$ can now be trivially carried out.
This is summarized by the general formula
\begin{eqnarray}
&& \int \frac{d^2 q}{(2 \pi)^2} (2 \pi) \delta (k + k_1 - |\bk+\bq| - |\bk_1 - \bq|)
F(\bq ,\bk, \bk_1) \nonumber \\
&&~~~~~~~~~~=\int_0^{2\pi} \frac{d\theta}{2\pi}
\int_0^{\infty} d\mu \frac{|\bk + \bk_1|^2}{4}
\left( \cosh^2 \mu - \cos^2 \theta \right) \delta(k+k_1 - |\bk +\bk_1| \cosh\mu)
F(Q,K,K_1) \nonumber \\
&&~~~~~~~~~~=\int_0^{2\pi} \frac{d\theta}{2\pi}
 \frac{|\bk + \bk_1| 
\left( \cosh^2 \mu - \cos^2 \theta \right)}{4 \sinh \mu} 
F(Q,K,K_1),
\label{thetaint}
\end{eqnarray}
where $F$ is an arbitrary function, and 
in the last equation we use the value of $\mu$ given in (\ref{muval})
and $Q$ given in (\ref{Qval}).

We are now prepared to discuss the handling of the infrared divergences
in (\ref{trans2}).
There are two distinct types of potential divergences and we will
consider them in turn.
\begin{itemize}
\item
The first singularity is associated with the divergence in the forward
scattering cross-section. By an examination of (\ref{deft12}) we see that
there is a $1/q^2$ divergence in $|T_1|^2$ when the momentum transfer $\bq$ approaches
zero, and also a $1/(\bq - \bk_1 + \bk)^2$
divergence in $|T_2|^2$ when the momentum transfer $\bq$ approaches 
$\bq = \bk_1 -\bk$; the latter divergence 
corresponds to forward scattering with an
exchange of particles. (Note that there is no divergence when
the momentum transfer $\bk + \bk_1$ vanishes, as is easily seen by an
examination of the $\lambda$
labels in (\ref{deft12}), combined with the expression (\ref{deft})).
However, we then notice from the right hand side of (\ref{trans2})
that the coefficient of $|T_1|^2$ in the collision term
vanishes at $\bq=0$; the term in the co-efficient
linear in $\bq$ will vanish after
an appropriate angular integral over $\bq$, and the remaining combination is
explicitly finite at $\bq = 0$.
Similar considerations apply to the 
the coefficient of $|T_2|^2$, which vanishes at $\bq=\bk_1 - \bk$.
The physics of this cancellation is simple and quite familiar. Its origin
is the ``$( 1 - \cos (\mbox{scattering angle}) )$'' phase space
factor that appears
in the expression for transport relaxation rates~\cite{am}: this vanishes
at zero scattering angle, and cancels out the divergence in the
scattering cross-section.

Now there remains the 
issue of how to numerically treat this potential divergence. It is clear
that the angular integral around $\bq=0$ and $\bq = \bk_1 - \bk$
should be taken with great care, or we will be left with a spurious divergence.
Here our elliptical parametrization (\ref{Qval}) turns out to be helpful.
Notice that, as must be the case, the points
$\bq = 0$ and $\bq = \bk_1 - \bk$ 
satisfy the delta function in (\ref{defellipse}), and therefore lie
on the ellipse defined by (\ref{muval}). These points lie at $\theta = \theta_0$
and $\theta = \theta_0 + \pi$ where $\theta_0$ is determined by the equation
\begin{equation}
2 \frac{K-K_1}{K+K_1} = e^{\mu + i \theta_0} + e^{-\mu-i\theta_0}
\end{equation}
This is a quadratic equation for $e^{i\theta_0}$, and therefore has two solutions
which are easily determined. For $\mu$ give by (\ref{muval}), one of the solution
will have unit modulus and therefore determines a real $\theta_0$: this is
guaranteed by the fact that $\bq=0$ lies on the ellipse defined by (\ref{muval}).
The angular averaging is carried out by the $\theta$ integral in (\ref{thetaint}),
and we now simply have to insure that the vicinities of the points $\theta = \theta_0$
(and $\theta = \theta_0 + \pi$) are treated symmetrically. More precisely, we always
pair the integrands at $\theta_0 + \theta$ and $\theta_0 - \theta$ ( and 
also at $\theta_0 + \pi + \theta$ and $\theta_0 + \pi - \theta$ ) so that
the combined integrand is explicitly finite at the potentially singular points.
\item
The second potential singularity arises purely from phase space considerations,
and is a curious property of energy-conserving
scattering of particles in two dimensions with the linear
dispersion $\varepsilon_k = k$.
We are considering the scattering of two incoming particles with momenta
$\bk$ and $\bk_1$. Consider the case where these two momenta
are collinear and point in the same direction {\em i.e.\/}
when $\bk \cdot \bk_1 = k k_1$. We are eventually integrating
over $\bk_1$ in the collision term of (\ref{trans2}), and this condition defines
a semi-infinite line in the two-dimensional $\bk_1$ space. Along this line,
we see from (\ref{muval}) that $\mu = 0$. Now the phase space factor
$1/\sinh \mu$ in (\ref{thetaint}) looks dangerous (generically the $\bq$ integral
in (\ref{thetaint}) will be non-singular, and will not effect the considerations
here). This infinite factor is present along the entire semi-infinite line
in the $\bk_1$ plane: if we pick a point a distance $y$ from this line
(almost collinear particles) we find that the $1/\sinh \mu$ factor diverges
as $1/y$. This implies a logarithmic divergence in the $\bk_1$ integral.
Fortunately, the specific form of $T_1$, $T_2$ under consideration here
comes to the rescue. It is easy to show from (\ref{deft},\ref{deft12})
that these cross-sections vanish identically for the case of the scattering
of particles moving in the same direction.
\end{itemize} 

The remaining steps in the solution of the integral equation were relatively
straightforward. It is easy to show that $g (k, \omega) \sim e^{-k/T}$ for large
$k$, and so it is an excellent approximation to simply truncate the infinite
range of $k$ integration at some large positive $k$. Over this finite
range, we expanded $g(k , \omega ) e^{k/T}$ over the space of Chebyshev polynomials.
We determined the action of the kernel of the integral equation on each Chebyshev
polynomials, and decomposed the resultant also over Chebyshev polynomials. This
defined a discrete matrix, which is the transcription of the kernel
to the discrete basis space of Chebyshev polynomials. Solving the integral equation
then simply became a matter of inverting the matrix. Notice that as the kernel
does not explicitly depend upon frequency, the same matrix could be used at any frequency.

\begin{figure}
\epsfxsize=5.5in
\centerline{\epsffile{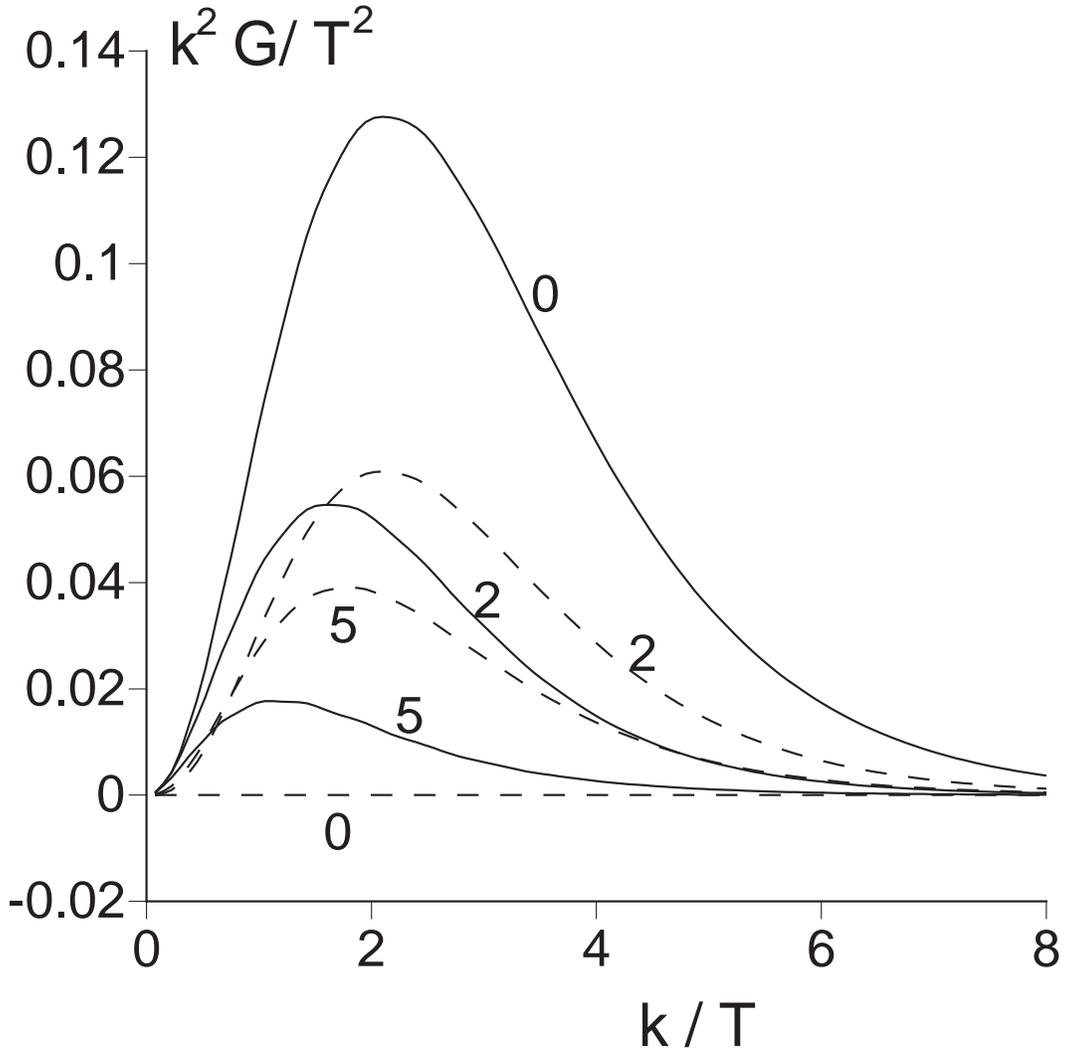}}
\vspace{0.2in}
\caption{Real and imaginary parts of the solution for the 
universal function $k^2 G/T^2$, introduced in (\protect\ref{scaleg}),
as a function of $k/T$ for a few values of 
$\omega/ \alpha^2 T$. The combination $k^2 G/T^2$ is 
that appearing in the integrand of the integral (\protect\ref{intscale})
for the conductivity.}
\label{fig3}
\end{figure}

We now present the results of the above numerical analysis. 
A simple dimensional analysis of (\ref{trans2}) shows that the entire $T$ and $\alpha$
dependence of  the integral equation can be scaled out, and the resulting
scaled equation is parameter free. In particular the solution of (\ref{trans2})
has the form
\begin{equation}
g(k, \omega ) = \frac{1}{\alpha^2 T^3} G \left( \frac{k}{T}, \frac{\omega}{\alpha^2 T}
\right),
\label{scaleg}
\end{equation}
where $G$ is a universal function we have determined numerically. Inserting the form
(\ref{scaleg}) into (\ref{paramet}) and (\ref{defj1}) we see that the conductivity
(\ref{eqp},\ref{eqp1}) is now replaced by
\begin{equation}
\widetilde{\sigma}^{{\rm qp}}_{xx} ( \omega )  = \frac{q^2 e^2}{\alpha^2 h}
\widetilde{\Sigma}^{{\rm qp}}_{xx} \left( \frac{\omega}{\alpha^2 T} \right)
\label{scalesqp}
\end{equation}
with the universal scaling function $\widetilde{\Sigma}^{{\rm qp}}_{xx}$ given by
\begin{equation}
\widetilde{\Sigma}^{{\rm qp}}_{xx} 
( \widetilde{\omega} ) = \int_0^{\infty} \bar{k}^2 
d \bar{k} G ( \bar{k} , \widetilde{\omega}) 
\label{intscale}
\end{equation}
This scaling function describes the broadening of the delta function in the real part
of (\ref{eqp1}): the large frequency behavior of (\ref{trans2}) shows that it
has the same total spectral weight as (\ref{eqp1}):
\begin{equation}
\int_0^{\infty} \frac{d \widetilde{\omega}}{\pi} {\rm Re} 
\left[\widetilde{\Sigma}^{{\rm qp}}_{xx} 
( \widetilde{\omega} ) \right] = \frac{\ln(2)}{2}
\end{equation}
We show the numerical solutions for the universal functions $G$ and 
$\widetilde{\Sigma}_{xx}^{{\rm qp}}$
in Figs~\ref{fig3} and~\ref{fig4} respectively.
The value of $\widetilde{\Sigma}_{xx}^{{\rm qp}} (0)$ determines the d.c. conductivity
\begin{equation}
\widetilde{\sigma}_{xx} ( \omega/T \rightarrow 0) = \left(\frac{q^2 e^2}{h} \right) 
\frac{0.437}{\alpha^2}
\label{lim1}
\end{equation}
This result and (\ref{sxxlarge}) and (\ref{sxylim}) completely specify the
limiting forms of the conductivity $\widetilde{\sigma}$.

\section{Synthesis}
\label{synth}

\begin{figure}
\epsfxsize=5.5in
\centerline{\epsffile{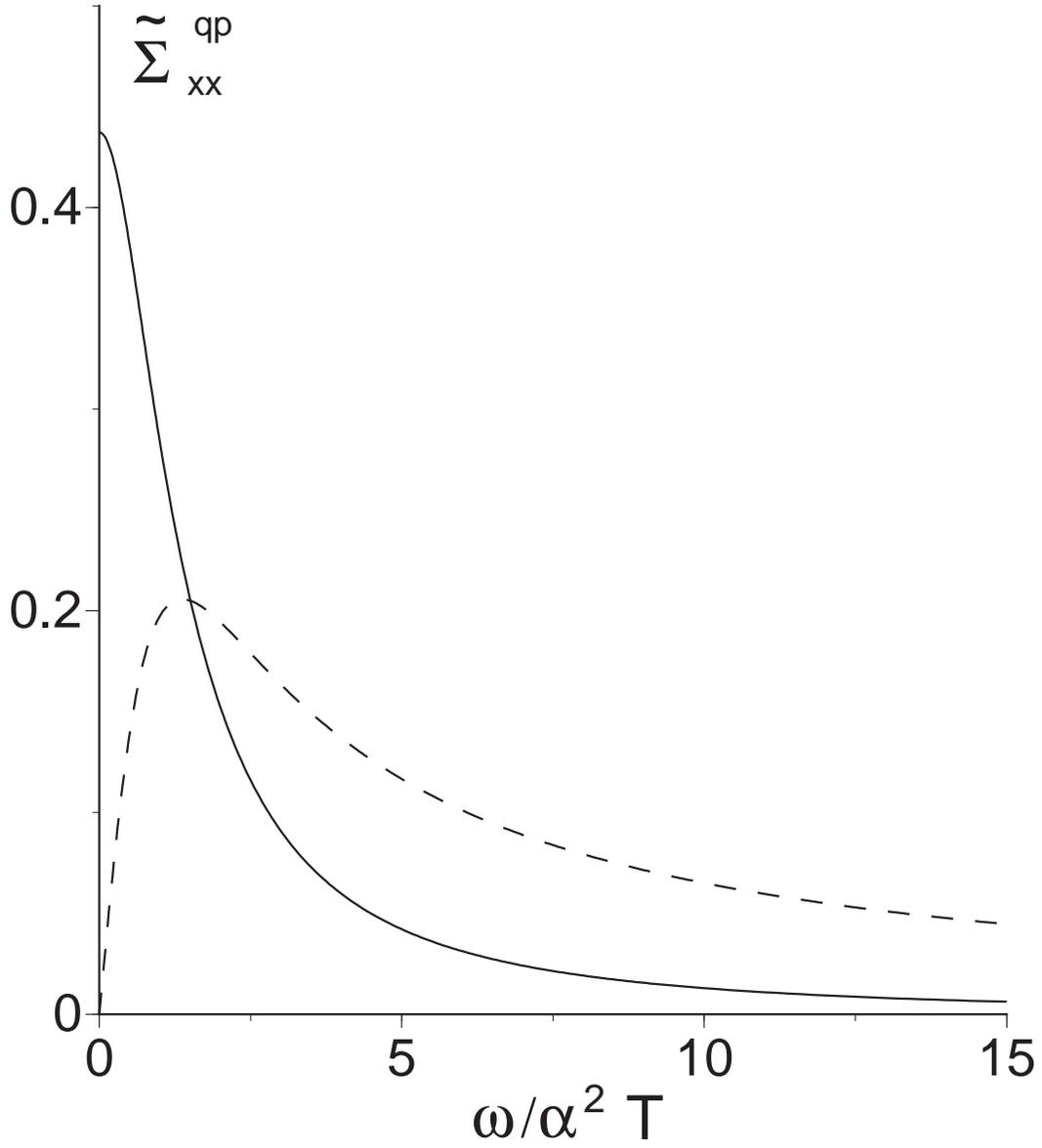}}
\vspace{0.2in}
\caption{Real and imaginary parts of the universal function $\widetilde{\Sigma}_{xx}^{{\rm qp}}$
as a function of $\omega / \alpha^2 T$. This result is related to 
$\widetilde{\sigma}_{xx}^{{\rm qp}}$ by (\protect\ref{scalesqp}).
}
\label{fig4}
\end{figure}

We will now finally turn to a discussion of the properties of the physical conductivity
$\sigma$. The several disparate ingredients necessary for its computation have all been 
assembled, and let us now summarize them. 
We began with a model with two species of anyons $\psi_{1,2}$. We defined their
irreducible conductivities $\widetilde{\sigma}_{1,2}$ and argued that the
physical conductivity $\sigma$ was related to them by (\ref{cs11}).
The $\psi_2$ anyons were non-critical and the value of $\widetilde{\sigma}_2$
is given simply by (\ref{cs12}). Sections~\ref{kubo} and~\ref{qtrans} were devoted
to studying the value of the critical $\widetilde{\sigma}_1$. 
Section~\ref{kubo} considered the perturbative
contribution which
in the limit $\omega/T \rightarrow \infty$ reduced to the coherent transport of
externally created particle-hole pairs. The Hall conductivity was given
entirely by such a contribution: $\widetilde{\sigma}_{xy}^{(1)}
+ \widetilde{\sigma}_{xy}^{(2)} + \widetilde{\sigma}_{xy}^{(3)}$ where the 
results are given in (\ref{sxy1}), (\ref{sxy2}) and (\ref{sxy3}) respectively.
The longitudinal conductivity consists of the coherent contribution
$\widetilde{\sigma}_{xx}^{{\rm coh}}$ given in (\ref{ecoh}), and an incoherent
quasiparticle contribution $\widetilde{\sigma}_{xx}^{{\rm qp}}$
which was computed in Section~\ref{qtrans};
the latter satisfies the scaling form (\ref{scalesqp}) and the scaling function
$\widetilde{\Sigma}_{xx}^{{\rm qp}}$ was plotted in Fig~\ref{fig4}.

The above discussion can be restated as follows. Let us define the dimensionless functions
$A$, $B$ by
\begin{eqnarray}
A(\omega ) &=& \frac{h}{q^2 e^2} \left(
\widetilde{\sigma}_{xx}^{{\rm qp}} (\omega)  + \widetilde{\sigma}_{xx}^{{\rm coh}}
(\omega ) \right) \nonumber \\
B(\omega ) &=& -\frac{1}{2} + \frac{h}{q^2 e^2} \left(
\widetilde{\sigma}_{xy}^{(1)} (\omega) + \widetilde{\sigma}_{xy}^{(2)} (\omega)
 + \widetilde{\sigma}_{xy}^{(3)} (\omega) \right)
\label{syn1}
\end{eqnarray}
Then, the physical conductivities are given by
\begin{eqnarray}
\sigma_{xx} (\omega) &=& \left(\frac{q^2 e^2}{h}\right) \frac{ A(\omega)}{\alpha^2 A^2 ( \omega) 
+ (
1 + \alpha B(\omega))^2}
\nonumber \\
\sigma_{xy} (\omega ) &=& \left(
\frac{q^2 e^2}{h}\right) \frac{ B(\omega) + \alpha( A^2 ( \omega ) + 
B^2(\omega))}{\alpha^2 A^2 ( \omega) + (
1 + \alpha B(\omega))^2}
\label{syn2}
\end{eqnarray}
We emphasize that $A$ and $B$ are complex, universal functions of $\omega/T$, and it is
necessary to include both their real and imaginary parts in the above expressions.

In evaluating our final results for $\sigma_{xx}$, an important complicating 
factor becomes apparent. While we are ultimately interested in $\alpha$ of order unity,
the terms entering (\ref{syn2}) are of many differing orders in $\alpha$ for small $\alpha$.
As a result, certain processes become unphysically dominant at small $\alpha$,
while in the real system we expect a more even-handed competition. Further the terms
also vary on differing scales of $\omega$. In particular, the quasiparticle/quasihole
conductivity $\widetilde{\sigma}_{xx}^{{\rm qp}}$ is of overall order
$1/\alpha^2$ and varies on a frequency scale $\omega \sim \alpha^2 T$; for small
$\omega$ it is the most dominant term as $\alpha \rightarrow 0$.
Of the other terms, $\widetilde{\sigma}_{xx}^{{\rm coh}}$ is order unity,
while $\widetilde{\sigma}_{xy}^{(1,2,3)}$ are of order $\alpha$,
and all vary on a scale $\omega \sim T$; these become important for
large $\omega$ as $\alpha \rightarrow 0$.
Because of these complications, we believe more significance should be attached to the form
of our
results for the individual conductivities in Section~\ref{kubo} and~\ref{qtrans}.
The final results, when these distinct components are combined, are expected to be
less reliable for moderate $\alpha$.

\begin{figure}
\epsfxsize=5.5in
\centerline{\epsffile{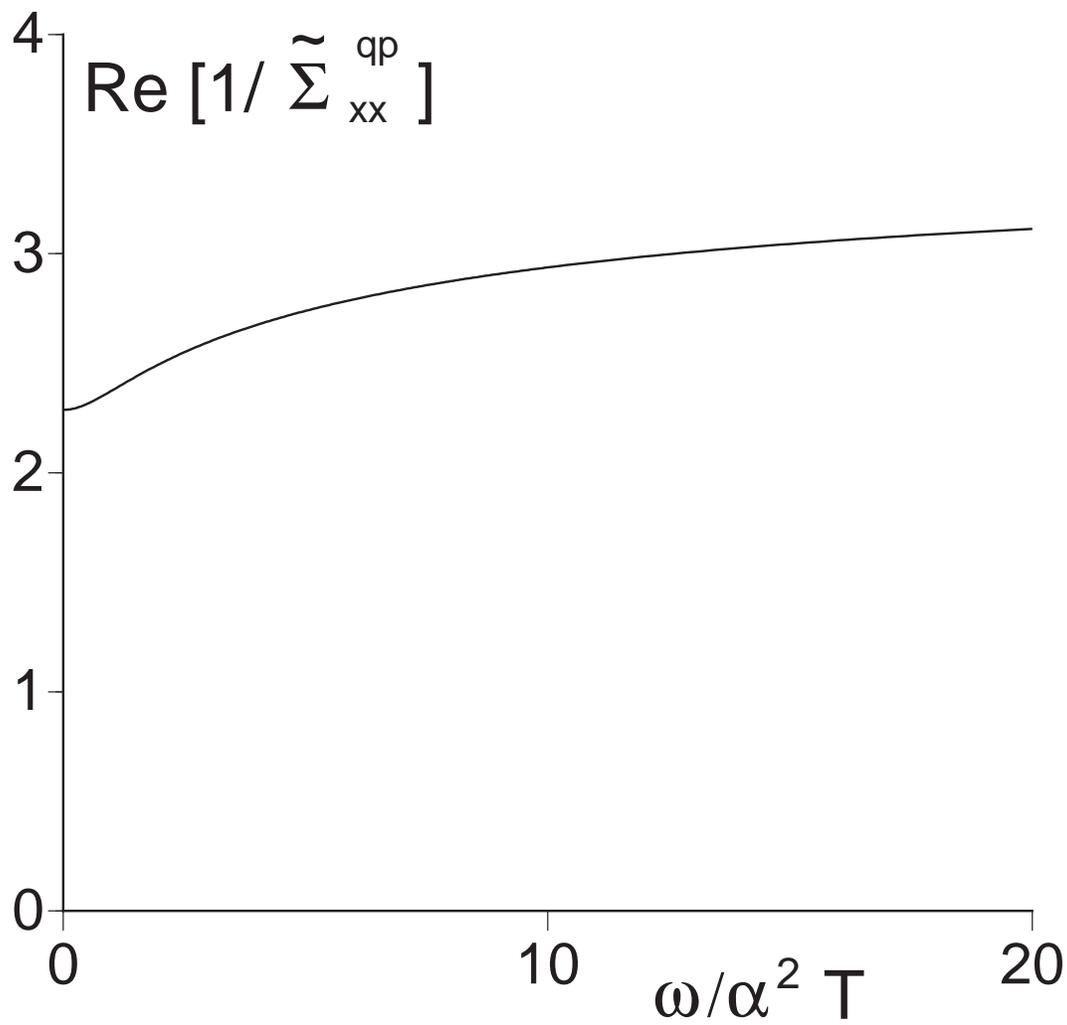}}
\vspace{0.2in}
\caption{Real part of $1/\widetilde{\Sigma}_{xx}^{{\rm qp}}$
as a function of $\omega / \alpha^2 T$.
}
\label{fig5}
\end{figure}

We begin by describing the formal structure of the small $\alpha$ expansion
for $\sigma$. It is necessary to consider two distinct frequency regimes separately:
\newline
(A) \underline{$\omega$ of order $\alpha^2 T$, but with $ \omega \ll T$}:
In this $\sigma_{xx}^{{\rm qp}}$ is the dominant term, and examination of the
small $\alpha$ limit of (\ref{syn2}) gives
\begin{equation}
\sigma_{xx} (\omega ) = \left(\frac{q^2 e^2}{h}\right) \left[
\frac{1}{\widetilde{\Sigma}_{xx}^{{\rm qp}} 
( \omega/ \alpha^2 T )} + \ldots \right] 
\label{syn3}
\end{equation}
We show a plot of the result above for ${\rm Re} \left[ \sigma_{xx} ( \omega ) \right]$ in 
Fig~\ref{fig5}. Notice that this not simply the inverse of the real plot
in Fig~\ref{fig4}, as the first equation in (\ref{syn3}) involves the inverse of the {\em 
complex} function $\widetilde{\Sigma}_{xx}^{{\rm qp}}$. The function 
${\rm Re}\left[ 1/\widetilde{\Sigma}_{xx}^{{\rm qp}} ( \widetilde{\omega} ) \right]$ 
approaches a finite
number in both the limits $\widetilde{\omega} \rightarrow 0$ 
(where it is simply the inverse of (\ref{lim1}) and $\widetilde{\omega}
\rightarrow \infty$:
\begin{eqnarray}
{\rm Re} \left[
\frac{1}{\widetilde{\Sigma}_{xx}^{{\rm qp}} (\widetilde{\omega} \rightarrow 0)}
\right] &=&  2.29 \nonumber \\
{\rm Re} \left[
\frac{1}{\widetilde{\Sigma}_{xx}^{{\rm qp}} (\widetilde{\omega} \rightarrow \infty)}
\right] &=& 3.47 
\end{eqnarray}
In a similar manner, we can also get the Hall conductivity in this regime:
\begin{equation}
\sigma_{xy} (\omega ) = \left(\frac{q^2 e^2}{h}\right) \left[
\frac{1}{\alpha} + \ldots \right]
\label{syn4}
\end{equation}
Surprisingly,
this result for $\sigma_{xy}$ has the opposite sign from the
Hall conductivity of the quantized Hall phase connected to the critical point under
consideration here: we strongly suspect this feature is an artifact of the absence 
scattering from interactions other than the Chern Simons term: a model in which
$\widetilde{\sigma}_{xx}^{{\rm qp}}$ is not so singular in the small $\alpha$ limit
should have a Hall conductivity of the ``correct'' sign. 
\newline
(B) \underline{$ \omega$ of order or larger than $T$}: Now the result is constructed simply
by inserting the perturbative results of Section~\ref{kubo}
for $\widetilde{\sigma}$ in (\ref{syn2}). The formal small $\alpha$ expansion is
\begin{eqnarray}
\sigma_{xx} ( \omega ) &=& (1 + \alpha) \widetilde{\sigma}_{xx}^{{\rm coh}} ( \omega )
 \nonumber \\
\sigma_{xy} ( \omega) &=& -\frac{q^2 e^2}{2h} + \widetilde{\sigma}_{xy} + \left[
-\frac{q^2 e^2}{4h} + \frac{h}{q^2 e^2}
\{\widetilde{\sigma}_{xx}^{{\rm coh}} ( \omega ) \}^2 \right]\alpha + \ldots,
\end{eqnarray}
and the ingredients determining the coefficients were plotted in Figs~\ref{fig1}
and~\ref{fig2}.

\begin{figure}
\epsfxsize=5.5in
\centerline{\epsffile{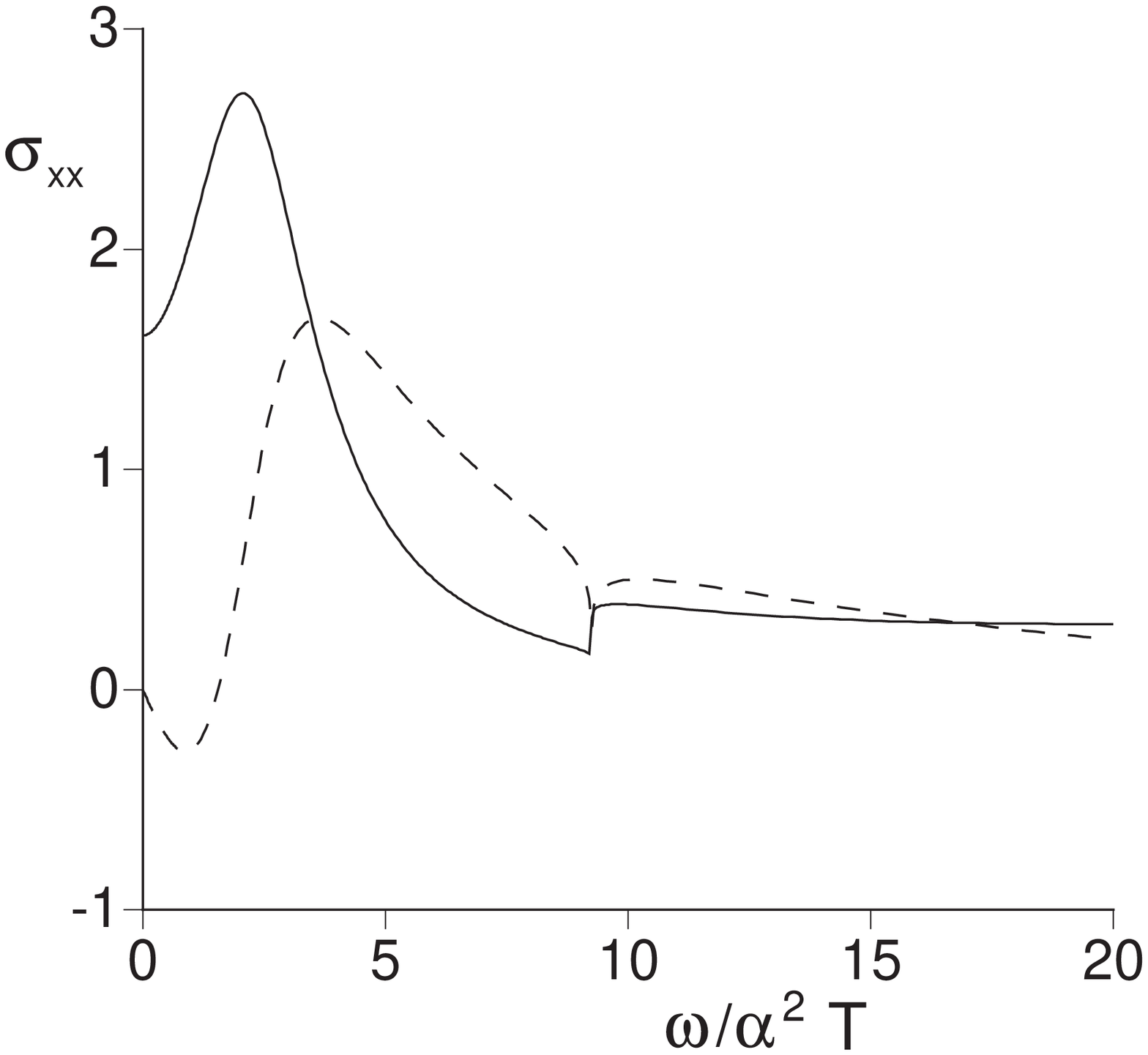}}
\vspace{0.2in}
\caption{Real and imaginary parts of the $\sigma_{xx}$ as a function of $\omega /\alpha^2 T$,
plotted at $\alpha = 0.3$. The conductivity is measured in units
of $q^2 e^2/h$. The singularities at $\omega = 2 M(T)$ ($\omega / \alpha^2 T
\approx 9.2$) will be rounded out at higher orders.
}
\label{fig6}
\end{figure}

\begin{figure}
\epsfxsize=5.5in
\centerline{\epsffile{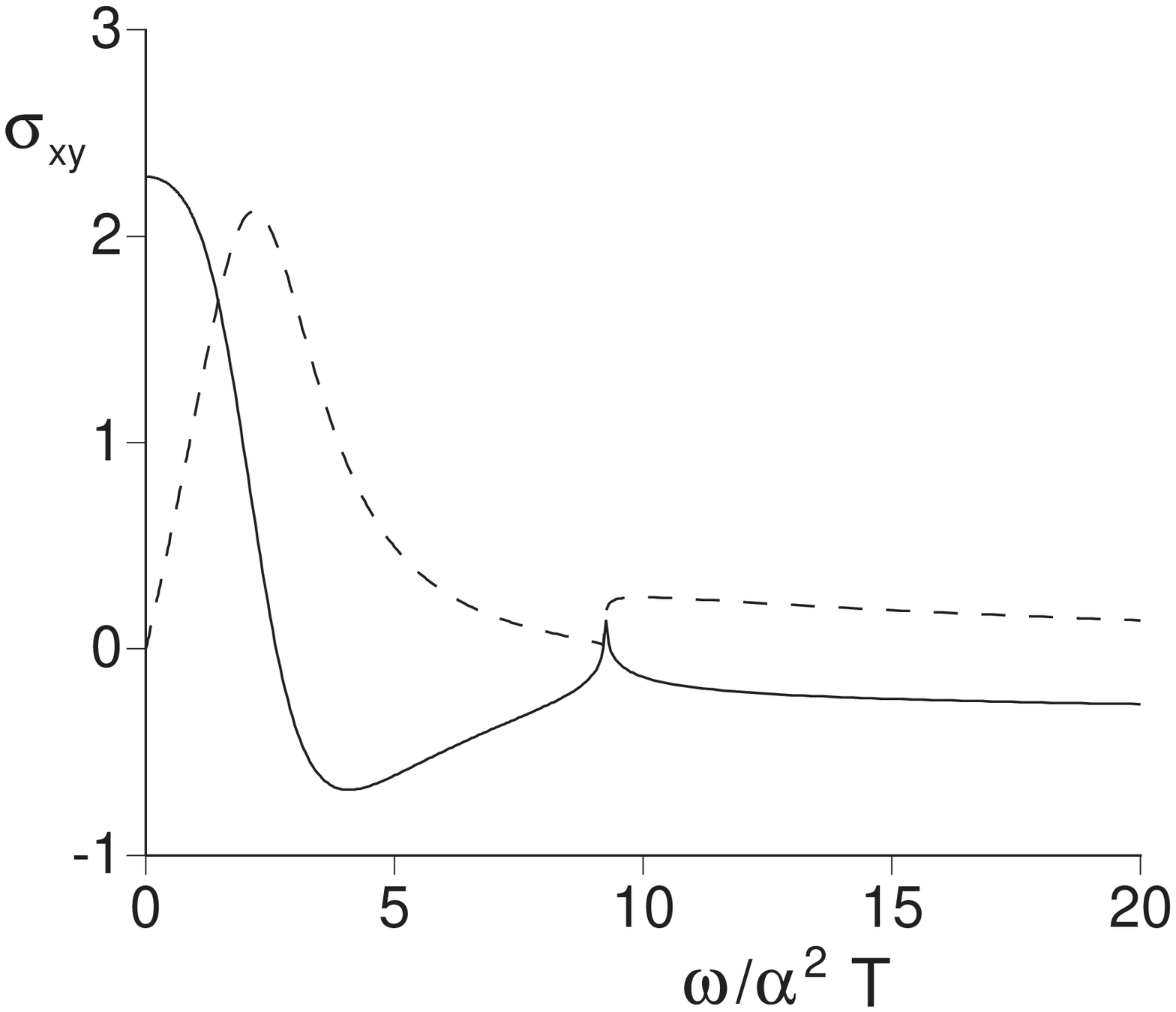}}
\vspace{0.2in}
\caption{Real and imaginary parts of the $\sigma_{xy}$ as a function of $\omega /\alpha^2 T$,
plotted at $\alpha = 0.3$. Again, the conductivity is measured in units
of $q^2 e^2/h$ and 
the singularities at $\omega = 2 M(T)$ ($\omega / \alpha^2 T
\approx 9.2$) will be rounded out at higher orders.
}
\label{fig7}
\end{figure}

An alternative approach to gaining some understanding of the predictions of
(\ref{syn2}) is to simply evaluate the full expression at a given value of $\alpha$.
This has been done in  
Figs~\ref{fig6} and~\ref{fig7}, which show
the results of directly evaluating (\ref{syn2})
at face value at $\alpha = 0.3$.

\section{Conclusions}
\label{conc}

The purpose of this paper was to describe the dynamic transport properties of the
the simplest possible quantum critical point in a quantum Hall system. In particular,
motivated by experimental results discussed in Section~\ref{intro}, we chose a model
quantum phase transition which had non-zero interactions in the critical fixed point theory.
The critical properties of such theories have a {\em universal\/} dependence of all
observables on $\omega /T$, and, as pointed in I, the dependence of the conductivity on
$\omega /T$ at 
two dimensional quantum critical points is especially important. 
The model we chose, ${\cal S}$ in (\ref{cs1}), is equivalent to a model of
free anyons and anti-anyons
with a relativistic Dirac spectrum. This model is properly considered as
an interacting theory for the following reason: two anyonic/anti-anyonic excitatios
have a non-trivial $T$ matrix for scattering~\cite{wilczek} in which the momenta of the
ingoing and outgoing particles can be arbitrary, and are subject only to the constraints
of conservation of total energy and momentum (it is clear that this fails only for 
purely bosonic or fermionic statistics, when the ingoing and outgoing momenta are the same). 
Such scattering is the main ingredient in 
establishing local equilibrium, and is essential to the 
description of transport in the regime $\omega \ll T$. 

Our results for the non-zero temperature transport properties were summarized in 
Section~\ref{synth}. It is evident from the discussion there that, while the general
scaling forms for the conductivities are very similar to those discussed in I for the
boson superfluid-insulator transition, the explicit results for the scaling functions
themselves are quite different. For instance, we found in I
that the longitudinal conductivity
$\sigma_{xx} (\omega) $ was a decreasing function of increasing $\omega$ near $\omega =0$.  
In the present quantum Hall problem, we found that for small $\alpha$, the longitudinal 
conductivity satisfied (\ref{syn3}), and is therefore seen to be an increasing function
of $\omega$ near $\omega = 0$, as shown in Fig~\ref{fig5}. 
This property was subsequently responsible for a peak in $\sigma_{xx}$ near
$\omega \sim T$ shown in Fig~\ref{fig6}.

As they stand,
our model calculations can probably not be directly
applied to microwave conductivity measurements in
the quantum Hall system~\cite{engel}. The most important defect that must be repaired
is the absence of Coulomb interactions among the charged excitations. A simple power counting
argument shows that the Coulomb interaction is a marginal perturbation at the fixed
point considered here. A theory for the consequences of this marginal perturbation
is not available at present, and appears to us to be
an important subject for future research.
It is interesting to note that since the Coulomb interaction is marginal at tree level,
it must be characterized by a dimensionless coupling constant. It is easy to see
that this coupling constant is
\begin{equation}
\frac{e^2}{\hbar v} = \left( \frac{e^2}{\hbar c} \right) \frac{c}{v} 
\approx \frac{c/v}{137}
\end{equation}
where $v$ is the velocity of the Dirac fermions, $c$ is the velocity of light,
and the quantity in the parenthesis is the dimensionless fine structure constant.
For $v$, it seems reasonable to use a typical drift velocity of electrons
in the two-dimensional electron gas $\sim 10^4 - 10^5 m/s$, and we see then
that this dimensionless coupling can be quite large. The Dirac fermion model (at $\alpha = 0$)
also describes quasiparticles in a d-wave superconductor, and so
the dimensionless coupling characterizing the Coulomb interactions can
also be quite large for the high temperature superconductors.
 
\acknowledgements
I am grateful to K.~Damle, S.~Girvin, N.~Read, R.~Shankar, S.~Sondhi, Y.-S.~Wu,
and J.~Ye for helpful discussions.
This research was supported by the National Science Foundation 
Grant No. DMR-96-23181. 

\appendix

\section{Large $N$ expansion}
\label{largeN}
Our study of transport near the superfluid-insulator quantum critical point in I,
and of the quantum Hall critical point in the present paper, has been carried
out using transport equations in which a four-point coupling between the charge
carriers is used as an expansion parameter. In this appendix we will outline an
alternative approach to the same problems, the large $N$ expansion. We shall not
provide an explicit numerical solution of the large $N$ transport equations here,
but it will evident that the structure of the results is very similar to those
already obtained.

We will begin with the simpler case of the superfluid-insulator transition
in a model of $O(2)$ quantum rotors which was discussed in I. We generalize this
to a model of $O(N)$ quantum rotors, and the quantum field theory describing the
critical point has the action
\begin{equation}
{\cal S}_r = \int_0^{1/T} d\tau \int d^2 x \left\{
\frac{1}{2} \left[ ( \partial_{\tau} \phi_{a} )^2 + c^2 ( \nabla_x \phi_{a}
)^2 +  m_{0c}^2  \phi_{a}^2 \right] 
+ \frac{u_0}{2N} ( \phi_{a}^2 )^2 \right\}.
\label{actionr}
\end{equation}
Here $\phi_{a}$ is a $N$-component field and the action has $O(N)$ symmetry
(the $O(N)$ index $a$ is implicitly summer over). 
The spatial and temporal gradient terms are both second order, so the action
has a ``Lorentz'' invariance with $c$ the velocity of light, and as a result
the dynamic critical exponent $z=1$; we will henceforth use units in which $c=1$.
The quartic non-linearity, $u_0$, has been scaled by a factor of $1/N$ to allow
the interesting large $N$ limit.
The bare ``mass'' term, $m_{0c}^2$, is cutoff dependent, and must be adjusted
order by order in $u_0$ so that the system is at its quantum critical point at $T=0$.
The superfluid-insulator transition of boson Hubbard models
is described by the case $N=2$, where $\Psi = \phi_1 + i \phi_2$ is the usual
complex superfluid order parameter. The case
$N=3$ which applies to quantum-critical points in quantum antiferromagnets~\cite{s,csy}.

First we review the large $1/N$ expansion for the order parameter susceptibility,
which is the propagator, $G$, of the $\phi_{a}$ field. This was extensively discussed in 
Ref~\cite{csy}. At momentum $\bk$, and Matsubara frequency $\omega_n$, the $N=\infty$
result for this propagator is
\begin{equation}
G(\bk, i \omega_n) = \frac{1}{\omega_n^2 + k^2 + \Theta^2 T^2},
\end{equation} 
where $\Theta$ is a pure (universal) number:
\begin{equation}
\Theta = 2 \ln \left( \frac{\sqrt{5} + 1}{2} \right)
\end{equation}
The $1/N$ corrections are obtained by decoupling the quartic term in (\ref{actionr})
by a Hubbard-Stratanovish field $\widetilde{\lambda}$ (not to be confused
with the charge subscript $\lambda$ used in Section~\ref{qtrans}, and also below).
There is then a three-point coupling $\sim \phi_a^2 \widetilde{\lambda}$ in the
action.  The propagator of the $\widetilde{\lambda}$ field is given by the 
inverse of the bubble graph of the $\phi_a$ field. In the low-energy, long-distance limit 
relevant for the critical properties, this propagator becomes independent of the
microscopic value of $u_0$, and equals
\begin{equation}
\frac{2}{N} \frac{1}{\Pi ( \bq, i\epsilon_n)}
\label{app1}
\end{equation}
where
\begin{equation}
\Pi ( \bq , i\epsilon_n ) = T \sum_{\omega_n} \int \frac{d^2 k}{(2 \pi)^2}
G(\bk, i\omega_n) G(\bk + \bq, i\omega_n + i\epsilon_n)
\end{equation} 
Notice that there is a factor of $1/N$ associated with each $\widetilde{\lambda}$
propagator: this comes from the sum over the $N$ components of the $\phi_a$ field
in the bubble graph. So the $1/N$ expansion for the susceptibility
is generated by expanding in the number of $\widetilde{\lambda}$ propagators.

Now we turn to transport properties. The value of the conductivity at $N=\infty$
was discussed in Section II of I. The structure was very similar to that obtained
in an expansion in $\epsilon=3-d$: there was a quasiparticle/quasihole delta function
at $\omega = 0$, and a continuum contribution at frequencies greater than
$\omega = 2 \Theta T$. The main purpose of this appendix is to determine the manner
in which the $\omega = 0$ delta function broadens out in the $1/N$ expansion.
Naturally, this will be done by formulating the transport equations of
the quasiparticles and quasiholes in the large $N$ limit.

The general $N$ transport equations for ${\cal S}_r$
were discussed in Appendix B of I in an expansion in $u_0$. Following that
treatment, we consider here the linear response to an ``electric'' field, ${\bf E}$,  which 
couples only
to the $\phi_{1,2}$ components {\em i.e.\/} the corresponding electric potential
couples to a generator of $O(N)$ which rotates the system in the $1,2$ plane.
Then it was argued that the distribution function of excitations
in the remaining components remained unchanged to linear order in the field---all corrections
are at least quadratic in ${\bf E}$. In other words, if, as in I, we make the mode expansion
in terms of particle creation and annihilation operators
\begin{equation}
\phi_{a} (\bx,t) = \int \frac{d^2 k}{(2 \pi)^2}
\frac{1}{\sqrt{2 \varepsilon_k}}
\left( b_{a} (\bk, t) e^{i \bk \cdot \bx} + b_{a}^{\dagger} (\bk, t) e^{- i
\bk \cdot \bx} \right),
\end{equation}
with the dispersion 
\begin{equation}
\varepsilon_k = (k^2 + \Theta^2 T^2)^{1/2},
\end{equation}
then, we have
\begin{equation}
f_a (\bk, t) \equiv \left\langle
b_a^{\dagger} ( \bk, t ) b_a (\bk, t) \right\rangle = n(\varepsilon_k) 
\equiv \frac{1}{e^{\varepsilon_k /T} - 1}~~~~\mbox{for $a > 2$}.
\end{equation}
This has very important consequences for the large $N$ expansion. Recall that the 
inverse propagator of the $\widetilde{\lambda}$ field was given by the sum
over the bubble graphs of the $N$ components of the $\phi_a$. In the large
$N$ limit, this sum is dominated by the $a>2$ components, and so also
remains unaffected to linear order in ${\bf E}$. So we can assume that the propagator
for $\widetilde{\lambda}$ is still given by (\ref{app1}).

Now the quantum transport equations can be easily written down by noticing that
the present problem has a structure closely analogous to the electron-phonon
problem~\cite{am,pk}: the $\widetilde{\lambda}$ field plays a role similar to the
phonons, and the $\phi_a^2 \widetilde{\lambda}$ coupling is analogous to the electron-phonon
coupling. The fact that the $\widetilde{\lambda}$ propagator is not modified
by ${\bf E}$ is equivalent to the statement that ``phonon drag'' effects are negligible
in the large $N$ limit. Indeed, the particular form of the propagator in (\ref{app1})
translates into a special phonon density of states, and the transport equations can
then be read directly off Ref~\onlinecite{pk}.
Let us define
\begin{equation}
b_{\pm} (\bk , t) \equiv \frac{b_1 (\bk, t) \pm b_2 (\bk, t)}{\sqrt{2}}.
\end{equation}
Then $f_{\lambda} ( \bk, t) = \left\langle b_{\lambda}^{\dagger} ( \bk, t)
b_{\lambda} ( \bk , t) \right \rangle$, with $\lambda = \pm$ 
satisfies the transport equation
\begin{eqnarray}
\left( \frac{\partial}{\partial t} + \lambda Q {\bf E} \cdot \frac{\partial}{\partial \bk}
\right) &&
f_{\lambda} ( \bk, t) =  -\frac{2}{N} \int_0^{\infty} \frac{d \Omega}{\pi}
\int \frac{d^2 q}{(2 \pi)^2}
{\rm Im} \left( \frac{1}{\Pi ( \bq, \Omega)} \right)  \nonumber \\
&& \times \Biggl\{
 \frac{(2 \pi) \delta ( \varepsilon_k - \varepsilon_{|\bk + \bq|} - 
\Omega )}{4 \varepsilon_k
\varepsilon_{|\bk+\bq|}}
 \biggl[ f_{\lambda} ( \bk , t) (1 + f_{\lambda} ( \bk + \bq, t))
(1 + n( \Omega )) \nonumber \\
&&~~~~~~~~~~~~~~~~~~~~~~~~~~~~-f_{\lambda} ( \bk+\bq , t) (1 + f_{\lambda} ( \bk, t))
n( \Omega )\biggr] \nonumber \\
&&~~~+ \frac{(2 \pi) \delta ( \varepsilon_k - \varepsilon_{|\bk + \bq|} + 
\Omega )}{4 \varepsilon_k
\varepsilon_{|\bk+\bq|}}
 \biggl[ f_{\lambda} ( \bk , t) (1 + f_{\lambda} ( \bk + \bq, t))
n( \Omega ) \nonumber \\
&&~~~~~~~~~~~~~~~~~~~~~~~~~~~~-f_{\lambda} ( \bk+\bq , t) (1 + f_{\lambda} ( \bk, t))
( 1 + n( \Omega ))\biggr] \nonumber \\
&&~~~+ \frac{(2 \pi) \delta ( \varepsilon_k + \varepsilon_{|-\bk + \bq|} -
\Omega )}{4 \varepsilon_k
\varepsilon_{|-\bk+\bq|}} \biggl[ f_{\lambda} ( \bk , t) f_{-\lambda} ( -\bk + \bq, t)
(1+ n( \Omega )) \nonumber \\
&&~~~~~~~~~~~~~~~~~~~~~~~~~~~~-(1+f_{-\lambda} ( -\bk+\bq , t)) (1 + f_{\lambda} ( \bk, t))
n( \Omega )\biggr] \Biggr\},
\label{app2}
\end{eqnarray}
where $Q$ is the charge of the $b_a$ quanta.
Now this equation can be linearized like (\ref{trans1}), and then solved using the numerical
methods discussed in Section~\ref{qtrans}. We will not present the numerical solution
here, but an essential fact is already clear from the structure of (\ref{app2}). 
The $\omega = 0$ delta function in the conductivity, found at $N=\infty$, gets broadened
out to a width $\omega \sim T/N$, and the peak conductivity at $\omega =0 $
is of order $N$.

The above large $N$ analysis can obviously be carried out at any $1<d<3$. It is useful
to do this and then take the limit of small $3-d$. We find that the equation so obtained
agrees {\em precisely} with the large $N$ limit of the small $3-d$ equation discussed
in Appendix B of I for general $N$. Thus the $N \rightarrow \infty$ and $3-d 
\rightarrow 0 $ limits of the quantum transport equations commute, and this is an important
check of the consistency of our analysis.

Our results can be easily extended to the large $N$ limit of 
either bosonic~\cite{ww} or fermionic~\cite{cfw}
theories with Chern-Simons terms. For instance, an alternative formulation of the theory
of anyons considered in this paper is to represent them as relativistic bosons
with attached statistical flux tubes: this is represented by a generalization 
of the action ${\cal S}_r$ as follows~\cite{ww}. Consider the case with $N = 2M$ even,
parametrize the $2N$ real fields $\phi_a$ in terms of $M$ complex fields $\Psi$
\begin{equation}
\Psi_a =  (\phi_{2a-1} + i  \phi_{2a})/\sqrt{2}~~~~~~~~~a =1\ldots M
\end{equation}
Now couple a $U(1)$ Chern Simons field to the diagonal $U(1)$ charge
\begin{eqnarray}
{\cal S}_{ra} = \int_0^{1/T} d\tau \int d^2 x && \Bigl\{ 
 | (\partial_{\tau} - i a_{\tau}) \Psi_{a} |^2 + 
c^2 | (\nabla_x - i a_x ) \Psi_{a} |^2
 +  m_{0c}^2  |\Psi_{a}|^2 
\nonumber \\
&&~~~~~~~\left. + \frac{u_0}{M} ( |\Psi_{a}|^2 )^2
+ \frac{iM}{4 \pi \alpha} \epsilon_{\mu\nu\lambda} a_{\mu}
\partial_{\nu} a_{\lambda} \right\}.
\label{actionra}
\end{eqnarray}
Notice that the effective statistcs parameter of the anyons is
$\alpha/M$: this is necessarily very close of the bosonic limit
for large $M$.

The large $M$ analysis of ${\cal S}_{ra}$ is very similar to that
of ${\cal S}$: in addition to the field $\widetilde{\lambda}$, we now
have a gauge field $a_{\mu}$ whose propagator is also given by a
boson current polarization bubble, and is of order $1/M$~\cite{ww}.
By arguments very similar to those for $\widetilde{\lambda}$, we can conclude
that the deviations in the $a_{\mu}$ propagator due to the external
electric field can be neglected {\em i.e.} ``photon drag'' is negligible
in the large $M$ limit.
The resulting quantum transport equation is then very similar
(\ref{app2}): along with the terms already displayed, we have an
additional parallel set with ${\rm Im} (1/\Pi )$ replaced
by the spectral density of the $a_{\mu}$ propagator times 
current matrix elements of the $\Psi$ field.
 
Finally, similar considerations apply to the large $M$ generalization of the
fermionic Chern-Simons theory, ${\cal S}$, of this paper.
We replicate the fermionic terms in ${\cal S}$ $M$ times, rescale $\alpha \rightarrow 
\alpha/M$, and then take the large $M$ limit as discussed above for the bosonic theories.

\section{Computations for the Hall conductivity}
\label{zeta}

This appendix shows how the result (\ref{sxy1n}) is obtained from (\ref{sxy1}).
First it is clear after using (\ref{self5}) and rescaling the integrand by $T$ that
\begin{equation}
\widetilde{\sigma}_{xy}^{(1)} ( \omega /T \rightarrow 0) = 
\frac{q^2 e^2 \alpha}{h} \ln (2) \int_{2 \alpha \ln(2)}^{\infty}
\frac{d \varepsilon}{\varepsilon^2} \tanh(\varepsilon/2)
\label{zeta1}
\end{equation}
Let us define
\begin{equation}
F(y) = \int_y^{\infty} \frac{dx}{x^2} \tanh (x/2),
\label{zeta2}
\end{equation}
and it is clear that to understand the small $\alpha$ limit of 
(\ref{zeta1}), we need the form of $F (y \rightarrow 0)$.
By integrating twice by parts, it is easy to show that
\begin{equation}
F(y) = \frac{\tanh(y/2)}{y} - \frac{\ln(y/2)}{2 \cosh^2 (y/2)}
+ \int_{y/2}^{\infty} dx \frac{ \ln(x) \sinh(x)}{\cosh^3 (x)}
\end{equation}
Now we can explicitly take the $y \rightarrow 0$ limit to get
\begin{equation}
F( y \rightarrow 0) = \frac{1}{2} \ln(1/y) + \frac{\ln(2) + 1}{2}
+ \int_0^{\infty} dx \frac{ \ln(x) \sinh(x)}{\cosh^3 (x)}
\label{zeta3}
\end{equation}
Then, we evaluate the last integral by first expanding the integrand
as an infinite series in $e^{-2x}$:
\begin{eqnarray}
\int_0^{\infty} dx \frac{ \ln(x) \sinh(x)}{\cosh^3 (x)}
&=& - 4 \int_0^{\infty} dx \ln(x) \left[
\sum_{n=1}^{\infty} (-1)^n n^2 e^{-2 n x} \right] \nonumber \\
&=& 2 \sum_{n=1}^{\infty} (-1)^n n ( \gamma + \ln(2 n) ) \nonumber \\
&=& 2 \left. \sum_{n=1}^{\infty} (-1)^n \left( ( \gamma + \ln(2) ) n^{1+s} 
+ \frac{d}{ds} n^{1+s} \right) \right|_{s=0} \nonumber \\
&=& 6 ( \gamma + \ln(2) ) \zeta(-1) + \left. 2\frac{d}{ds} \left(
(2^{2+s} - 1) \zeta(-1-s) \right) \right|_{s=0} \nonumber \\
&=& -\frac{\gamma}{2} - \frac{7 \ln(2)}{6} - 6 \zeta^{\prime} (-1).
\label{zeta4}
\end{eqnarray}
Notice that the intermediate steps involve badly divergent series in $n$, 
and so the validity of the final result is not a priori obvious. 
To check it, we numerically evaluated the left hand side and verified that it was equal to
the final right hand side to a very high accuracy.
Combining (\ref{zeta1}), (\ref{zeta2}), (\ref{zeta3}) and (\ref{zeta3}),
we can now obtain (\ref{sxy1n}).

\end{document}